\documentclass[epj,final]{svjour}
\usepackage{dsfont}
\usepackage{graphicx,amsmath}

\def\be{\begin{eqnarray}}
\def\ee{\end{eqnarray}}
\def\bc{\begin{center}}
\def\ec{\end{center}}

\def\rmF{{\rm F}}
\def\rmd{{\rm d}}
\def\om{\omega}

\def\half{{\textstyle \frac12}}
\def\quart{{\textstyle \frac14}}
\newcommand{\lsim}{\stackrel{\scriptstyle <}{\phantom{}_{\sim}}}
\newcommand{\gsim}{\stackrel{\scriptstyle >}{\phantom{}_{\sim}}}
\begin{document}
\title{
Scalar quanta in Fermi liquids: zero sounds, instabilities, Bose condensation, and
a metastable state in dilute nuclear matter
 }
\author{
E.\ E.\ Kolomeitsev\inst{1} and D.\ N.\ Voskresensky\inst{2}
}
\authorrunning{E.E. Kolomeitsev and D.N. Voskresensky}
\titlerunning{Scalar quanta in Fermi liquids}
\institute{
\inst{1}Matej Bel University, SK-97401 Banska Bystrica, Slovakia\\
\inst{2}National Research Nuclear University (MEPhI), 115409 Moscow, Russia
}
\date{}
\abstract{
The spectrum of bosonic scalar-mode
excitations in a normal Fermi liquid with  local scalar interaction is investigated for various values and momentum
dependence of the scalar Landau parameter $f_0$ in the
particle-hole channel. For $f_0 >0$ the conditions are found when
the phase velocity on the spectrum of  zero sound acquires a
minimum at  non-zero momentum. For $-1<f_0 <0$ there are only
damped excitations, and for $f_0<-1$ the spectrum becomes unstable
against the growth of scalar-mode excitations. An effective Lagrangian for the scalar excitation
modes is derived after performing a bosonization procedure. We
demonstrate that the  instability may be tamed by the
formation of a static Bose condensate of the scalar modes. The
condensation may occur in a homogeneous or inhomogeneous state
relying on the momentum dependence of the scalar Landau parameter.
We show that in the isospin-symmetric nuclear matter there may appear a metastable state at subsaturation nuclear density owing to the condensate.
Then we consider a possibility of the condensation of the
zero-sound-like excitations in a state with a non-zero momentum in
Fermi liquids moving with overcritical velocities, provided an
appropriate momentum dependence of the Landau parameter $f_0
(k)>0$. We also argue that in peripheral heavy-ion collisions the Pomeranchuk instability may occur already for $f_0 >-1$.
}

\maketitle
\section{Introduction}

The theory of normal Fermi liquids was built up by
Landau in Ref.~\cite{FL}, see in textbooks~\cite{LP1981,Nozieres,GP-FL}.
The Fermi liquid approach to the description of nuclear systems
was developed by Migdal in Refs.~\cite{Mjump-1,Mjump-2,Mjump-3,M67,M67a}. In the Fermi liquid
theory the low-lying excitations are described by several
phenomenological Landau parameters. Pomeranchuk has shown in
Ref.~\cite{Pomeranchuk} that Fermi liquids are stable only if some
inequalities on the values of the Landau parameters are fulfilled.

In this work we study low-lying scalar excitation modes
(density-density fluctuations) in  cold normal Fermi liquids
for various values and momentum behavior of the scalar Landau
parameter $f_0$ in the particle-hole channel.  We assume that an
interaction in the particle-particle channel is repulsive and the
system is, therefore, stable against pairing in an s-wave state.
An induced p-wave pairing to be possible at very low temperatures
$T<T_{{\rm c},p}$, see  Ref.~\cite{pwave-pairing-1,pwave-pairing-2,pwave-pairing-3}, can be
precluded by the assumption that the temperature of the system is
small but above $T_{{\rm c},p}$.

For $f_0 > 0$ in a Fermi liquid there exist zero-sound excitations. In some cases the momentum dependence of the parameter $f_0$  can be such that the phase velocity
at the zero-sound branch possesses a minimum at a non-zero momentum. This
means that the spectrum satisfies the Landau necessary condition
for superfluidity. As a consequence this may lead to a
condensation of zero-sound-like excitations with a non-zero momentum in
moving Fermi liquids with the velocity above the Landau critical
velocity~\cite{Vexp95}. Similar phenomena may occur in moving
He-II, cold atomic gases, and other moving media, like rotating
neutron stars, cf.
Refs.~\cite{Pitaev84,V93,Melnikovsky,BP12,KV2015,Kolomeitsev:2015dua}.
For $-1<f_0 <0$ excitations are damped, for $f_0<-1$ the spectrum
is unstable against the growth of zero-sound-like modes and
hydrodynamic modes. Up to now it was thought that for $f_0<-1$ the
mechanical stability condition is violated that results in
exponential buildup of the density fluctuations. In hydrodynamic
terms the condition $f_0<-1$ implies that the speed of the first
sound becomes imaginary. In a one-component Fermi liquid this would lead to an exponential growth of the aerosol-like mixture of droplets and bubbles (spinodal instability) resulting in a mixed liquid-gas like stationary state. In the isospin-symmetric nuclear matter  (if the Coulomb interaction is neglected) the liquid-gas phase transition might occur~\cite{RMS,SVB,Chomaz:2003dz} for low temperatures ($T\lsim 20$ MeV) and the baryon densities $n\lsim 0.7 n_0$, where $n_0$ is the nuclear saturation density. In a many-component system a mechanical instability is accompanied by a chemical instability, see Ref.~\cite{Margueron:2002wk}. The inclusion of the Coulomb interaction, see Refs.~\cite{Ravenhall:1983uh,Maruyama:2005vb}, leads to a possibility of the pasta phase in the neutron star crusts for densities $0.3 n_0\lsim n\lsim 0.7 n_0$. For higher densities in dense neutron star interiors there may appear phase transitions to the pion, kaon, charged rho condensate states and to the quark matter \cite{Migdal78,MSTV90,Tatsumi:2002dq-1,Tatsumi:2002dq-2,MTVTEC,Voskresensky:1997ub-1,Voskresensky:1997ub-2}. In some models these phase transitions are considered as  first order phase transitions leading to a mixed  pasta phases.

In this work we study scalar excitations in Fermi liquids. The key point of this
work is that we suggest that at certain conditions  for $f_0<-1$ the Pomeranchuk instability may result in an accumulation of a static Bose condensate of a scalar field.
The condensate amplitude is stabilized by the repulsive
self-interaction.  The condensation may occur in the homogeneous
either in inhomogeneous state depending on the momentum dependence
of the Landau parameter $f_0$. In the presence of the
condensate the Fermi liquid proves to be stable. Also we consider condensation of
scalar excitations in moving Fermi liquids. Then we apply our results to the isospin-symmetric nuclear matter. We argue for the softening of the equation of state and occurrence of a new metastable state at a subsaturation nuclear density associated with the formation of the condensate state.

The work is organized as follows. In  Sect.~\ref{sec:sounds} we
study the spectrum of excitations in a one-component Fermi liquid in
the scalar channel in dependence on $f_0$. In Sect.~\ref{sec:boson} we bosonize the local interaction and suggest an effective Lagrangian for the self-interacting scalar
modes. In Sect.~\ref{sec:Bosecond} we study the Pomeranchuk instability
for $f_0<-1$ and discuss possibilities of homogeneous and inhomogeneous Bose condensate ($\phi$) of scalar modes. In Sect.~\ref{sec:pomer} we show that in the presence of the $\phi$ condensate the zero- and first-sound modes become stable and apply our results to the isospin-symmetric nuclear matter.  Dynamics of the system is also briefly discussed. In Sect.~\ref{sec:moving} we consider a condensation of scalar excitations in moving Fermi liquids with repulsive interactions, $f_0 (k)>0$. Concluding remarks are formulated in Sect.~\ref{sec:conclude}.

\section{Excitations in a Fermi liquid}\label{sec:sounds}

\subsection{Landau particle-hole amplitude.}

Consider the simplest case of a one-component Fermi liquid of
non-relativistic fermions.  As discussed in the Introduction, we
assume that the system is stable against pairing. The
particle-hole scattering amplitude on the Fermi surface obeys the
equation~\cite{LP1981,Nozieres,GP-FL,M67}
\begin{align}
&\widehat{T}_{\rm ph}(\vec{n}\,',\vec{n};q) =
\widehat{\Gamma}^{\om}(\vec{n}\,',\vec{n})
 \nonumber\\
&\quad+ \langle \widehat{\Gamma}^{\om}(\vec{n}\,',\vec{n}'')\,
\mathcal{L}_{\rm ph}(\vec{n}\,'';q)\, \widehat{T}_{\rm
ph}(\vec{n}\,'',\vec{n};q) \rangle_{\vec{n}\,''}\,,
\label{Tph-FL}
\end{align}
where $\vec{n}$ and $\vec{n}\,'$ are the directions of the fermion momenta
before and after scattering and $q=(\om,\vec{k})$ is the momentum
transferred in the particle-hole channel. The brackets stand for
averaging over the momentum direction $\vec{n}$
\begin{align}
\langle\dots\rangle_{\vec{n}} = \intop\frac{\rmd \Omega_{\vec
n}}{4\,\pi} \big(\dots\big)\,, \label{angl-aver}
\end{align}
and the particle-hole propagator is
\begin{align}
\mathcal{L}_{\rm ph}(\vec{n};q)= \intop_{-\infty}^{+\infty}\!\!
\frac{\rmd \epsilon}{2\pi i}\!\! \intop_{0}^{+\infty}\!\!\frac{
\rmd p\,p^2}{\pi^2} \, {G}(p_{\rmF+})\,{G}(p_{\rmF-})\,,
\end{align}
where we denoted $p_{\rmF
\pm}=(\epsilon\pm\om/2,p_\rmF\,\vec{n}\pm\vec{k}/2)$ and $p_{\rm
F}$ stands for the Fermi momentum. The quasiparticle contribution
to the full Green function is given by
\begin{align}
G(\epsilon,\vec{p})=\frac{a}{\epsilon-\xi_{\vec{p}}+i\, 0\, {\rm
sign} \epsilon}\,,\quad\xi_{\vec{p}}=\frac{p^2-p_\rmF^2}{2\,m_{\rm
F}^*}\,.\label{Gn-QP}
\end{align}
Here $m^*_{\rm F}$ is the effective fermion mass, and the parameter $a$ determines a quasiparticle
weight in the fermion spectral density, $0<a\leq 1$, which is expressed through the retarded
fermion self-energy $\Sigma_{ \rm F}^R(\epsilon,p)$ as $a^{-1}= 1-(\partial \Re\Sigma_{ \rm
F}^R/\partial \epsilon)_0$. The full Green function contains also a regular background part
$G_{\rm reg}$, which is encoded in the renormalized particle-hole interaction $\hat{\Gamma}^\om$ in Eq.~(\ref{Tph-FL}).

The interaction in the particle-hole channel can be written as
\begin{align}
\hat\Gamma^{\om}(\vec{n}\,',\vec{n}) =
\Gamma^{\om}_0(\vec{n}\,'\vec{n})\,  \sigma'_0 \sigma_0 +
\Gamma^{\om}_1(\vec{n}\,'\vec{n})\,
(\vec{\sigma}\,'\vec{\sigma})\,.
\label{Gom-fullspin}
\end{align}
 The matrices $\sigma_j$ with $j=0,\dots,3$ act on incoming
fermions while the matrices $\sigma'_j$ act on outgoing fermions;
$\sigma_0$ is the unity matrix and other Pauli matrices
$\sigma_{1,2,3}$ are normalized as ${\rm
Tr}\sigma_i\sigma_j=2\delta_{ij}$. We neglect here the spin-orbit
interaction, which is suppressed for small transferred momenta
$q\ll p_{\rm F}$. The scalar and spin amplitudes in
Eq.~(\ref{Gom-fullspin}) can be expressed in terms of
dimensionless  scalar and spin Landau parameters
\begin{align}
\tilde{f}(\vec{n}\,',\vec{n})
&=a^2\,N_0\Gamma_0^\om(\vec{n}\,',\vec{n})\,, \nonumber\\
\tilde{g}(\vec{n}\,',\vec{n}) &=a^2\,N_0
\Gamma_1^\om(\vec{n}\,',\vec{n})\,, \label{L-param}
\end{align}
where the normalization constant is chosen as in applications to
atomic nuclei~\cite{M67,SaperFayans,MSTV90} with the density of
states at the Fermi surface, $N_0=N(n=n_0)$, taken at the nuclear
saturation density $n_0$ and $N=\nu m_\rmF^*\,p_\rmF/\pi^2$ with $\nu =1$ for one type of fermions and $\nu =2$ for two types of fermions, like for the isospin-symmetric nuclear matter. Such a normalization is at variance with that used, e.g., in Refs.~\cite{FL,LP1981,Nozieres,GP-FL}. Their
parameters are related to ours defined in Eq.~(\ref{L-param}) as $f=N \,\tilde{f}/N_0$ and $g=N\,\tilde{g}/N_0$. The baryon density, $n$, and the Fermi momentum, $p_\rmF$, are related as $n=\nu\,p_\rmF^3/(3\pi^2)$.

The Landau parameters can be expanded in terms of the Legendre
polynomials $P_l(\vec{n}\cdot\vec{n}\,')$,
\begin{align}
\tilde{f}(\vec{n}\,',\vec{n})=\sum_l \tilde{f}_l\, P_l
(\vec{n}\cdot\vec{n}\,')\,,
\end{align}
a similar expression exists for the parameter $\tilde{g}$.

The Landau parameters ${f}_{0,1}$, ${g}_{0,1}$ or $\tilde{f}_{0,1}$, $\tilde{g}_{0,1}$ can be
directly related to observables~\cite{GP-FL}. For instance,
using the standard expression for the variation of the chemical potential with the particle density for $T=0$~\cite{LP1981,M67},
\begin{align}
\delta \mu_\rmF =\frac{\delta E_\rmF}{\delta n} =\frac{2\epsilon_\rmF}{3n} \delta n + \frac{f_0}{N}\,\delta n \,,
\label{muf}
\end{align}
one finds the incompressibility of the system
\begin{align}
K=n\,\frac{\delta\mu_\rmF}{\delta n}=\,\frac{p^2_{\rm F}}{3m_\rmF^*}\big(1+f_{0}\big)\,.
\label{comp}
\end{align}
Here $E_{\rm F}$ is the energy density of the Fermi liquid and $\epsilon_\rmF=p_\rmF^2/(2\,m_\rmF^*)$.

Similarly, the square of the first sound velocity  is expressed as
\begin{align}
u^2 =\frac{\partial P}{\partial \rho}=\frac{p_{\rm F}^2}{3m_{\rm F} m^{*}_{\rm F}}
(1+f_{0})\,,
\label{u}
\end{align}
$P=n\,\rmd E_\rmF/\rmd n -E_\rmF$ is the pressure, $\rho$ is the mass density.
The positiveness of the incompressibility and of the first-sound velocity squared are assured by
fulfillment of the Pomeranchuk condition $f_0>-1$. The effective quasiparticle mass is given by~\cite{LP1981,Nozieres}
\begin{align}
\label{mef}
\frac{m^*_{\rm F}}{m_{\rm F}} = 1+a^2\,N\,\overline{\Gamma_0(\cos\theta)\cos\theta}=
1+\frac13 f_1\,,
\end{align}
where the bar denotes the averaging over the azimuthal and polar
angles. The positiveness of the effective mass is assured by
fulfillment of the Pomeranchuk condition $f_1>-3$. Note that the
traditional normalization of the Landau parameters (\ref{L-param})
depends explicitly on the effective mass $m^*_{\rm F}$ through the density
of states $N$. Therefore, it is instructive to rewrite
Eq.~(\ref{mef}) using the definition in Eq.~(\ref{L-param})
\begin{align}
\frac{m^*_{\rm F}}{m_{\rm F}}=\frac{1}{1-\frac13
\frac{m_\rmF}{m^*_{\rmF}(n_0)}\tilde{f}_1}\,.
\end{align}
From this relation we obtain the constraint
$\tilde{f}_1\!<\!3\,m^*_{\rmF}(n_0)/$ $m_\rmF$ for the effective mass to
remain positive and finite; otherwise the effective mass tends to
infinity in the point where
$\tilde{f}_1=3\,m^*_{\rmF}(n_0)/m_\rmF$.
Thus, for the systems
where one expects a strong increase of the effective mass, the
normalization (\ref{L-param}) of the Landau parameters would be
preferable.
On the other hand, right from the
dispersion relation follows that\begin{align} \frac{m^*_{ \rm
F}}{m_{ \rm F}}=\frac{1-(\partial \Re\Sigma_{ \rm F}^R/\partial
\epsilon)_\rmF}{1+(\partial \Re\Sigma_{ \rm F}^R/\partial
\epsilon_p^0)_\rmF}\,,
\end{align}
where $\epsilon_p^0 =p^2/2m_{\rm F}$. Thereby, the Landau parameter
$\tilde{f}_1$ can be expressed via the energy-momentum derivatives
of the fermion self-energy.

Below we focus our study on effects associated with the zero harmonic
$\tilde{f}_0$ in the expansion of $\Gamma^{\om}_{0}$ as a function
of $(\vec{n}\vec{n}\,')$. Then the solution of Eq.~(\ref{Tph-FL})
is
\begin{eqnarray}
\widehat{T}_{\rm ph}(\vec{n}\,',\vec{n};q) &=& T_{{\rm ph},0}(q)\,
\sigma'_0\, \sigma_0 + T_{{\rm
ph},1}(q)\,(\vec{\sigma}\,'\vec{\sigma})\,, \nonumber\\ T_{{\rm
ph},0(1)}(q) &=& \frac{1}{1/\Gamma^\om_{0(1)}- \langle
\mathcal{L}_{\rm ph}(\vec{n};q)\rangle_{\vec{n}}}\,.
\label{Tph-sol}
\end{eqnarray}
The averaged particle-hole propagator is expressed through the Lindhard function
\begin{align}
\langle \mathcal{L}_{\rm ph}(\vec{n};q)\rangle_{\vec{n}}=-a^2 N
\Phi\big(\frac{\om}{v_\rmF\,k},\frac{k}{p_\rmF}\big)\,,
\end{align}
where
\begin{align}
\Phi(s,x)&= \frac{z_-^2-1}{4(z_+-z_-)}\ln\frac{z_- + 1}{z_- - 1}
\nonumber\\ & -\frac{z_+^2-1}{4(z_+-z_-)}\ln\frac{z_+ + 1}{z_+ -
1} +\frac12\,. \label{LindF}
\end{align}
Here and below we use the dimensionless variables $z_\pm= s \pm x/2$, $s=\om/kv_{\rm F}$, $x=k/p_{\rm F}$. For real $s$ the Lindhard function acquires an imaginary part
\begin{align}
\label{ImPhi}
\Im \Phi(s,x) =\left\{
\begin{array}{ccl}
 \frac{\pi}{2}\,s &,& 0\leq s\leq 1-\frac{x}{2}
\\
\frac{\pi}{4x}(1-z_{-}^2) &,&
1-\frac{x}{2}\leq  s\leq  1+\frac{x}{2}
\\
0 &,& \mbox{otherwise}
\end{array}
\right.
\,.
\end{align}
For $s \gg 1$,
\begin{align}
\Phi(s,x) \approx -1/(3\, z_+\, z_-)\,.
\end{align}
For $x\ll 1$ the function $\Phi$ can be expanded as
\begin{align}
\label{PhiLow-x}
\Phi (s, x)\approx
1 + \frac{s}{2} \log\frac{s-1}{s+1} -\frac{x^2}{12 \left(s^2-1\right)^2}\,,
\end{align}
and if we expand it further for $s\ll 1$ we get
\begin{align}
\label{PhiLow}
\Phi (s, x)\approx
1  +i \frac{\pi}{2}s -s^2 -\frac{x^2}{12} -\frac{s^4}{3}  -\frac{x^2s^2}{6} -\frac{x^4}{240}\,.
\end{align}

The amplitudes (\ref{Tph-sol}) possess simple poles and
logarithmic cuts. For the amplitude $T_{{\rm ph},0}$ the pole is
determined by the equation
\begin{eqnarray}
\frac{1}{f_0}
=-\Phi\big(\frac{\om}{v_\rmF\,k},\frac{k}{p_\rmF}\big)\,.
\label{sound-eq}
\end{eqnarray}
A similar equation exists in the $g$-channel (with replacement $f_0\to
g_0$). Analytical properties of the solution (\ref{sound-eq}) have
been studied in \cite{Sadovnikova-1,Sadovnikova-2,Sadovnikova-3}.

Expanding the retarded particle-hole amplitude $T^R_{{\rm
ph},0}(q)$ near the spectrum branch
\begin{align}
T^R_{{\rm ph},0}(q)&\approx \frac{2\om (k)V^2(k)}{(\om+i0)^2
-\om^2 (k)}\,, \nonumber \\ V^{-2}(k) & =a^2 N
\frac{\partial\Phi}{\partial \om}\Big|_{\om (k)}\,, \label{Tnr}
\end{align}
we identify the quantity
\begin{align}
D^R(\om,k)=[(\om+i0)^2 -\om^2 (k)]^{-1}\label{TnrD}
\end{align}
as the near-pole expansion of the retarded propagator of {\em a scalar  boson} with the dispersion relation $\om =\om(k)$ and the quantity $V(k)$ as the effective vertex of the fermion-boson interaction. This scalar boson can be associated with a field operator $\hat{\phi}$ in the second quantization scheme.

For the neutron matter  the parameters $f=f_{nn}$, $g=g_{nn}$ are the neutron-neutron Landau scalar and spin parameters. Generalization to the two-component system, e.g., to the nuclear
matter of arbitrary isotopic composition is formally simple~\cite{M67,M67a}. Then the amplitude should be provided  with four indices ($nn$, $pp$, $np$ and $pn$). However, equations for the partial amplitudes do not decouple. For the isospin-symmetric nuclear  system with the omitted Coulomb interaction the situation is simplified since then $f_{nn}=f_{pp}$ and $f_{np}=f_{pn}$. In this case one usually presents $\Gamma_0^\om$ in Eqs.~(\ref{Gom-fullspin}) and (\ref{L-param}) as $\Gamma_0^\om =(\tilde{f}+\tilde{f}'\vec{\tau}'\vec{\tau})/a^2N_0$, and
$\Gamma_1^\om =(\tilde{g}+\tilde{g}'\vec{\tau}'\vec{\tau})/a^2N_0$, where $\tau_i$ are isospin Pauli matrices.
The parameters can be extracted from the experimental data on atomic nuclei. Unfortunately, there are essential uncertainties in numerical values of some of
these parameters. These uncertainties are, mainly, due to attempts to get the best fit
to experimental data in each concrete case slightly modifying parametrization used
for the residual part of the $NN$ interaction.
For example, basing on the analysis of Refs.~\cite{KS80-1,KS80-2}, with the normalization $C_0 =1/(a^2 N_0)=300$\,MeV$\cdot$fm$^{3}$ one gets $\tilde{f}_0 \simeq 0.25$, $\tilde{f}_0' \simeq 0.95$, $\tilde{g}_0 \simeq 0.5$, $\tilde{g}_0' \simeq 1.0$, see Table 3 in~\cite{M67a}. The parameters $\tilde{g}_0 $, $\tilde{g}_0'$   are rather slightly density dependent whereas $\tilde{f}_0 $ and   $\tilde{f}_0'$ depend on the density essentially.  For $\tilde{f}_0 (n)$ Ref.~\cite{M67a} suggested to use the  linear density dependence, then with above given parameters we have  $\tilde{f}_0 (n)=-2.5+2.75\, n/n_0$.

Quantities $f$ and $f'$ (and similarly $g$ and $g'$) are expressed through $f_{nn}$ and $f_{np}$ as $f=\frac{1}{2}(f_{nn}+f_{np})$ and $f'=\frac{1}{2}(f_{nn}-f_{np})$.  The amplitudes of these four channels ($f$, $f'$, $g$, $g'$) decouple, cf. Ref.~\cite{AAB}. The
excitation modes are determined by four equations similar to Eq.~(\ref{sound-eq}), now with $f_0$, $f_0'$, $g_0$ and $g_0'$. For strongly isospin-asymmetric matter, e.g., for the neutron matter, there are no data from which the Landau parameters can be extracted. In this case the parameters are calculated within a chosen model for the $NN$ interaction, see review \cite{Backman:1984sx}.

Electromagnetic interaction leads to a coupling between density oscillations and charge oscillations. Thereby, the presence of the small Coulomb potential in finite size systems modifies the low-lying modes~\cite{AAB} determined by Eq.~(\ref{sound-eq}) (now with $f_0$, $f_0'$, $g_0$ and $g_0'$). A typical value of the frequency shift is $\delta\om \sim \overline{\delta V}\sim Ze^2/R$, where $Z$ is the charge and $R$ is the radius of the nucleus. Since in all cases which we are interested in  $\overline{\delta V}\ll\epsilon_{\rm F}$, we  ignore the Coulomb effects in our exploratory study.

At finite temperatures the Lindhard function $\Phi$ should be replaced by the
temperature-dependent Lindhard function $\Phi_T$ calculated in Ref.~\cite{Voskresensky:1982vd}, see also in review~\cite{MSTV90}. For that the zero-temperature Green functions entering the Lindhard function should be replaced to the temperature dependent ones. Generalization of the expansion~(\ref{PhiLow}) for low temperature case ($T\ll \epsilon_{\rm F}$) is then given by
\begin{align}
\label{PhiT}
\Phi_T (s, x, t)=  \Phi(s,x)\, \Big(1-\frac{\pi^2}{12} t^2\Big)\,,
\end{align}
where $t=T/\epsilon_{\rm F}$ and $\epsilon_{\rm F}=p_{\rm F}^2/2m_{\rm F}^{*}$ is the Fermi energy. For high temperatures $(T\gg\epsilon_{\rm F}$) we have
\begin{align}
\label{PhiT1}
\Phi_T (0,x,t) =\frac{2}{3\, t}
\Big( 1-\frac{x^2}{6\,t}-\frac{1}{3\,\sqrt{2\pi\,t^3}}\Big)\,.
\end{align}

We discuss now the properties of the bosonic modes for different values of the Landau parameter $f_0$.

\subsection{Repulsive interaction ($f_0>0$). Zero sound.}\label{sec:spec-f0pos}

For $f_0>0$  there exists a real solution of Eq.~(\ref{sound-eq})
such that for $k\to 0$ the ratio $\om_{s}(k)/k$ tends to a
constant. Such a solution is called the zero sound. The zero sound
exists as a quasi-particle mode in the high frequency limit $\om
\gg 1/\tau_{\rm col}$, where $\tau_{\rm col}\propto\epsilon_{\rm
F}/T^2$ is the fermion collision time. In the opposite limit it
turns into a hydrodynamic (first)
sound~\cite{Pethick-Ravenhall88}. At some value $k_{\rm lim}\lsim
p_\rmF$ the spectrum branch enters the region with $\Im\Phi> 0$,
and the zero sound becomes a damped diffusion mode.

We search the solution of Eq.~(\ref{sound-eq}) in the form
\begin{align}
\om_s(k)=k\, v_\rmF\, s(x),
\end{align}
where $s(x)$ is a function of $x=k/p_\rmF$ which we take in the form $s(x)\approx s_0 + s_2\,
x^2+s_4\,x^4$. The odd powers of $x$ are absent since $\Re\Phi$ is
an even function of $x$.

Above  we assumed that $f$ depends only on $\vec{n}\vec{n}'$. The
results are, however, also valid if the Landau parameter $f$ is a
very smooth function of $x^2$. From now we suppose
\begin{align}
\label{fexp}
f_0 (x)\approx f_{00} +f_{02} x^2\,,
\end{align}
where the parameter $f_{02}$ is determined by the effective range
of the fermion-fermion scattering amplitude and expansion is valid
provided $|f_{00}|\gg |f_{02}|$ for relevant values $x\lsim p_{\rm
F}$. According to Ref.~\cite{SaperFayans}, in the case of atomic
nuclei ($n\simeq n_0$), $f_{02}=- f_{00} r_{\rm eff}^2 p_{\rm
F}^2/2$, and  $0.5\lsim r_{\rm eff}\lsim 1$fm, as follows from the
comparison with the Skyrme parametrization of the nucleon-nucleon
interaction and with the experimental data. For isospin-symmetric
nuclear matter $f_{00}>0$ for $n\gsim n_0$, $f_{00}<0$ for lower
densities, and in a certain  density interval below $n_0$,
$f_{00}<-1$, cf. Refs.~\cite{M67a,SVB,Backman:1984sx,Speth:2014tja,Matsui,Maslov:2015wba}. In the purely neutron
matter one has $-1< f_{00} < 0$ for $n\lsim
n_0$~\cite{Wambach:1992ik}.

The constant term, $s_0$,  follows from Eq.~(\ref{sound-eq})
\begin{align}
\frac{1+f_{00}}{f_{00}}=\frac{s_0}{2}\ln\frac{s_0+1}{s_0-1}
\,. \label{s0-coeff}
\end{align}
For  $f_{00}\ll 1$ the solution of Eq.~(\ref{s0-coeff}) is
\begin{align}
s_{0}= & 1 + C \big[1 + (4+ 5 f_{00})(C/2 f_{00})
\nonumber\\ &+ (24 + 52 f_{00} + 29 f_{00}^{2})(C/2
 f_{00})^2\big]\, \,,
\end{align}
where $C=(2/e^2)\, \exp(-2/f_{00})$. For $f_{00}\le 1.8$ this expression reproduces the numerical solution with a percent precision. In the opposite limit $f_{00}\gg 1$ the asymptotic solution is $s_0 =\sqrt{f_{00}/3}$.
The  coefficients $s_2$ and $s_4$ follow as
\begin{align}
s_2 & = \left[\frac{f_{02}}{f_{00}^{2}} - \frac12 \frac{\partial^2
\Phi}{\partial x^2}\Big|_{s_0,0}\right] \Big[\frac{\partial
\Phi}{\partial s}\Big|_{s_0,0}\Big]^{-1} \label{s2-coeff}\\& =
\frac{ s_0 (\alpha +f_{02}) (s_0^2-1)}{f_{00}\,(1+ f_{00} -
s_0^2)}\,,\quad\alpha=f_{00}^2/[12(s_0^2-1)^2]\,. \nonumber
\end{align}
\begin{align}
s_4 &= s_0 f_{00}\frac{s_2^2 +
\frac{1 + 5 s_0^2-80(s_0^2-1) s_0 s_2}{240(s_0^2-1)^2}-\frac{f_{02}^2 }{f_{00}^{3}} (s_0^2-1)^2 } {(s_0^2-1) (1+f_{00}-s_0^2)}\,.
\label{s4-coeff}
\end{align}

\begin{figure}
\centerline{\includegraphics[width=7cm]{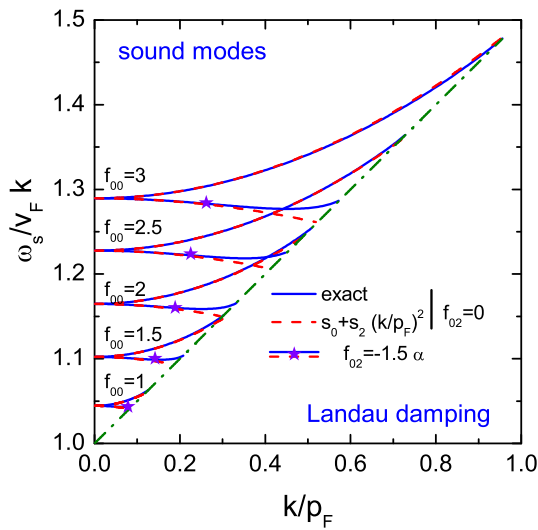}}
\caption{(Color online) Spectrum of the zero-sound modes defined by
Eq.~(\ref{sound-eq}) for various values of the Landau parameters
$f_{00}$ and $f_{02}$. Solid lines show the results of the
numerical solution, dashed lines are the quadratic expansions with
the parameters $s_0$ and $s_2$ given by Eqs.~(\ref{s0-coeff}) and
(\ref{s2-coeff}). The values of $f_{00}$ are shown in labels.
Lines with (without) stars are calculated for $f_{02}=0$
($f_{02}=-1.5\alpha$), where $\alpha$ is defined in Eq.
(\ref{s2-coeff}). The dash-dotted line shows the border of the
imaginary part of the Lindhard function (\ref{LindF}). Below
this line the frequency is complex.} \label{fig:omS}
\end{figure}

The numerical solutions of Eq.~(\ref{sound-eq}) are shown in
Fig.~\ref{fig:omS}  by solid lines for various values of the
Landau parameters $f_{00}$ and $f_{02}$.  The zero-sound
solution exists for $s_0>0$.  We check that the inequality
$1+f_{00}-s_0^2
>0$ holds for $f_{00}>0$.  The quadratic approximation for the spectrum  $\om\approx
v_\rmF\,k\,[s_0+s_2 x^2]$ is demonstrated by dashed lines.
 Only the quasi-particle part of the spectrum   is
shown with $\om_s(k)/(v_\rmF\, k)>  1+ k/(2p_\rmF)$. This
inequality holds for $k/p_{\rmF}<k_{\rm
max}/p_\rmF=(1-\sqrt{1-16s_2(s_0-1)})/(4\,s_2)$ if $s_2\le
1/[16(s_0-1)]$. For larger $k$ the Lindhard function becomes
complex for continuation of the branch $\om=\om_s(k)$ and the
solution of Eq.~(\ref{sound-eq}) acquires an imaginary part. For
$s_2>1/[16(s_0-1)]$,   the function $\om_s(k)$ does not enter in
the region of the complex Lindhard function. However, for positive
$s_2$ there is another source of the mode dissipation related to
the decay of one mode's quantum in two quanta. The latter process
is allowed if the energy-momentum relation $\om_s (p) =\om_s
(p')+\om_s (|\vec{p}-\vec{p}{\,'}|)$ holds, that is equivalent to
the relation $ 1-\cos\phi  = 3s_2/(s_0 p_{\rm F}^2)$\,, where
$\cos\phi =(\vec{p}\,\vec{p}{\,'})/(p\,p')$, which is fulfilled if
$s_2>0$.

In Fig.~\ref{fig:omS} we demonstrate first the case $f_0={\rm
const}$, i.e., $f_{02}=0$. As we see, in this case $s_2>0$ for any
$f_{00}>0$. We see that the quadratic approximation coincides very
well with the full solution. Then we study how the spectrum
changes if the parameter $f_{02}$ is taken nonzero. As we conclude
from Eq.~(\ref{s2-coeff}) the coefficient $s_2$ can be negative
for $f_{02}<-\alpha\,.$ The latter implies: $f_{02}\lsim -10$ for
$f_{00}=1$; $f_{02}\lsim -2.6$ for $f_{00}=2$; and $f_{02}\lsim
-1.7$ for $f_{00}=3$. In Fig.~\ref{fig:omS}  by solid curves and
dashed curves marked with stars we depict the zero-sound spectrum
for $f_{02}< -\alpha$, being  computed following
Eq.~(\ref{sound-eq})  and within the quadratic approximation,
respectively. We chose here $f_{02}=-1.5 \alpha$. The quadratic
approximation for $s(x)$ works now worse for momenta close to
$k_{\rm max}$ and the next term $s_4 x^4$ should be included. We
note that the parameter $s_4$ is positive in this case and the
function $s(x)$ has a minimum at $x_{\rm min}=\sqrt{|s_2|/(2
s_4)}$. In the point $k=k_{0}$ corresponding to the minimum of
$\om_s(k)/k$ the group velocity of the excitation $v_{\rm
gr}=d\om_s/dk$ coincides with the phase one $v_{\rm ph}=\om_s/k$.
The quantity $\om_s (k_0)/k_0$ coincides with the value of the
Landau critical velocity $u_{\rm L}$ for the production of Bose
excitations in the  superfluid moving with the velocity $u>u_{\rm
L}$.

\begin{figure}
\centerline{\includegraphics[width=7cm]{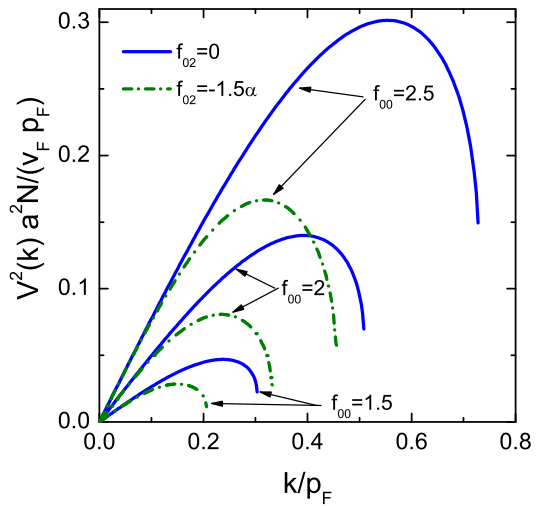}}
\caption{(Color online) Coupling constant of zero-sound modes and
fermions as a function of the momentum $k$ for various values of
the Landau parameters $f_{00}$ and $f_{02}$, calculated from
Eq.~(\ref{Tnr}). All curves end at the limiting value of $x_{\rm
max}=k_{\rm max}/p_\rmF$ at which the spectral branch enters into
the region of complexity of the $\Phi$ function. } \label{fig:g2k}
\end{figure}

The effective vertex of the boson-fermion coupling (\ref{Tnr}) is
shown in Fig.~\ref{fig:g2k} as a function of $x$ for several
values of $f_{00}$ and $f_{02}$. We plot it in the range
$0<x<x_{\rm max}=k_{\rm max}/p_\rmF$, where the quasiparticle
branch of the zero-sound spectrum is defined. For $f_{02}=0$ the
vertex $V(k)$ is shown by solid lines, for $f_{02}< 0$ by
dash-dotted lines. We see that $V^2 (x_{\rm max})$ is smaller for
$f_{02}< 0$ than for   $f_{02}=0$. For small $x$  one can use the
analytical expression
\begin{align}
&a^2 N \frac{V^2(x)}{v_\rmF\, p_\rmF} =
x\Big[\frac{\partial\Phi}{\partial s}\Big|_{s(x),x}\Big]^{-1}
 \nonumber\\
&\quad \approx \frac{s_2 xf_{00}^2}{\alpha+f_{02}}\Big[1+\frac{4\alpha s_2 x^2
}{\alpha+f_{02}}\Big(6  s_2-\frac{s_0}{s_0^2-1}\Big)\Big]\,,
\label{g2k-exp}
\end{align}
which works well for $k\lsim 0.75 k_{\rm max}$. For $k\to 0$ we
get $a^2 N {V^2(k)}\simeq  s_0 (s_0^2-1) f_{00}k
v_{\rmF}/(1+f_{00}-s_0^2)$.

\subsection{Moderate attraction, $-1<f_{00}<0$. Diffusons.}

Assume $f_{02}=0$. For $-1 < f_{00} < 0$, Eq.~(\ref{sound-eq}) has
only damped solutions with $\Re s<1 $ and $\Im s <0$.  In the
limit of $s,x\ll 1$, using the expansion (\ref{PhiLow}) we easily
find the analytic solution
\begin{align}
\label{omnegf}
\frac{\om_{\rm d}(k)}{kv_{\rm F}} =i s_{\rm d}(x) \approx -i\frac{2}{\pi} \left[\frac{1-|f_{00}|}{|f_{00}|}+\frac{x^2}{12}\right]\,,
\end{align}
which is valid for $1-|f_{00}|\ll 1$ and $x\ll 1$. We see that the
solution we found is purely imaginary and its dependence on $x$ is
very weak.

\begin{figure}
\centerline{\includegraphics[width=7cm]{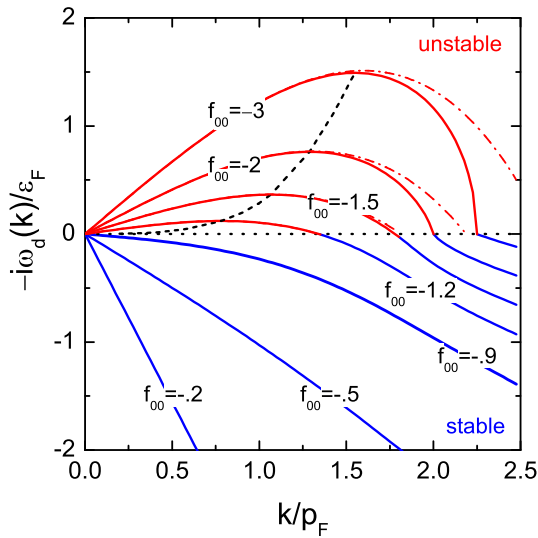}}
\caption{(Color online) Solid lines show the purely imaginary solutions of
Eq.~(\ref{eq-omD}) for various values of the Landau parameter
$f_{00} <0$. Dashed line indicates positions of the maxima of
spectra $\om_{\rm d}$ for $f_{00}<-1$. Dash-dotted lines show
the approximated spectra $\om_{\rm d}(k)\approx i v_\rmF k (s_{\rm
d0}+ s_{\rm d2}\,x^2+ s_{\rm d4} x^4)$ with the parameters from
Eqs.~(\ref{sd0}) and (\ref{sd24}). } \label{fig:omD}
\end{figure}

More generally, for purely imaginary $s=i\,\tilde{s}$,
$\tilde{s}\in R$, the Lindhard function can be rewritten as
\begin{align}
&\Phi(i\tilde{s},x) =\widetilde{\Phi}(\tilde{s},x)= \frac{1}{4x}\big[\tilde{s}^2-\quart
x^2+1\big] \log \frac{\tilde{s}^2+\left(\half
x+1\right)^2}{\tilde{s}^2+\left(\half x-1\right)^2}
\nonumber\\
&\quad + \frac{{\tilde{s}}}{2}\left[{\rm
arctan}\frac{\tilde{s}}{1-\half x} + {\rm
arctan}\frac{\tilde{s}}{1+\half
x}\right]+\frac{1-\pi\tilde{s}}{2}\,.
\label{Phist}
\end{align}
We note that the function ${\rm
arctan}(\tilde{s}/[1-\half x])$ should be continuously extended for $x>2$ as
${\rm arctan}(\tilde{s}/[1-\half x])+\pi{\rm sign}\tilde{s}$\,.
Solutions of equation
\begin{align}
1/|f_{00}|=\Phi(i\tilde{s},x)
\label{eq-omD}
\end{align}
for $0>f_{00}>-1$ are depicted in Fig.~\ref{fig:omD} for
$f_{00}=-0.2$, $-0.5$, $-0.9$. These solutions are damped ($-i\om
<0$). For $f_{02}\neq 0$ (at the condition $0>f_0 (k)>-1$),   the
damped character of the solutions does not change.

\subsection{Strong attraction, $f_{00}<-1$. Pomeranchuk instability.}\label{sec:pominstab}

Now we consider the case of a strong attraction in the scalar channel $f_{00}<-1$. We continue to assume $f_{02}=0$ for certainty.  The Pomeranchuk instability for $f_{00}<-1$ manifests
itself in the negative incompressibility, which follows from the relation (\ref{comp}) after the replacement $f_0\to f_{00}$. The incompressibility $K\propto 1 + f_{00}$, being negative for $f_{00} <-1$.

Solutions of Eq.~(\ref{eq-omD}) for $f_{00}<-1$  are  shown in
Fig.~\ref{fig:omD}. We see that there is an interval of $k$, $0 <
k < k_{\rm d}$, where $-i\om_{\rm d}>0$ and, hence, the mode is
exponentially growing with time. This corresponds to the
Pomeranchuk instability in a Fermi system with strong scalar
attraction, see Ref.~\cite{Nozieres}. For $k>k_{\rm d}$ we get
$-i\om_{\rm d}(k>k_{\rm d})<0$, and the mode becomes stable again.
The  momentum $k_{\rm d}$ is determined from Eq.~(\ref{eq-omD})
for $\tilde{s}=0$:
\begin{align}
\frac{1}{f_{00}}=\frac{x_{\rm d}^2-4}{8x_{\rm d}} \log
\left|\frac{x_{\rm d} + 2}{x_{\rm d} - 2} \right|- \frac{1}{2}\,,\label{omD-end}
\end{align}
$x_{\rm d}=k_{\rm d}/p_\rmF$.
The instability increment $-i\om_{\rm d}$ has a maximum at some momentum $k_m$. The locations of maxima for different $f_{00}$ values are connected by dashed line in Fig.~\ref{fig:omD}.

The spectrum of the unstable zero-sound-like mode can be written for $x<<1$ as
$s_{\rm d}(x)\approx s_{\rm d 0}+ s_{\rm d 2}x^2+s_{\rm d
4}\,x^4$, where the leading term is determined by the equation
\begin{align}
\frac{\pi}{2}-\arctan s_{\rm d 0}=\frac{z_f}{s_{\rm d0}}\,,\quad z_f=1-1/|f_{00}| \,.
\label{sd0}
\end{align}
The subleading terms are equal to
\begin{align}
s_{\rm d 2}&=\frac{s_{\rm d0}}{12(1+s_{\rm d0}^2)(z_f\, (1+s_{\rm d0}^2) -s_{\rm d0}^2)}\,,
\nonumber\\
s_{\rm d 4}&=\frac{1-5 s_{\rm d0}^2}{20(1+s_{\rm d0}^2)^2}s_{\rm d2}
- \frac{4 s_{\rm d0} s_{\rm d2}^2}{1+s_{\rm d 0}^2} -12 s_{\rm d2}^2\,.
\label{sd24}
\end{align}
This approximation for the spectrum is illustrated in Fig.~\ref{fig:omD} by dash-dotted lines. The approximation works very well for $-1.5<f_{00}<-1$ (dash-dotted lines and solid lines coincide) {\it but becomes worse for smaller $f_{00}$.}

For a slightly subcritical case, where $0<z_f\ll 1$, we obtain
\begin{align}
\label{omD-unst}
\om_{\rm d}(k) \approx i\frac{2}{\pi}v_{\rm F}k\,[z_f-k^2/(12p_\rmF^2)]\,.
\end{align}
The function $-i\om_{\rm d}$ has a maximum at $k_m =2p_\rmF\sqrt{z_f}$ equal to
\begin{align}
\label{maxsound}
-i\om_m= ({8}/{3\pi}) v_{\rm F}p_{\rm F}\, z_f^{3/2}\,.
\end{align}
For $T\neq 0$ corrections to these results are $1+O(T^2/\epsilon_{\rm F}^2)$.

The first-sound velocity squared follows from (\ref{u}) after the  replacement $f_0\to f_{00}$. For $f_{00}<-1$ as the incompressibility the first sound velocity $u^2\propto 1+f_{00}$ is negative. Thus, in the region $f_{00} (n)<-1$ the hydrodynamical first-sound mode with the frequency $\omega =u q$ for $q\to 0$ proves to be unstable. Note that the first sound exists in the hydrodynamical (collisional) regime, i.e. for $\om\ll \tau_{\rm col}^{-1}\propto T^2$,  which is the opposite limit  to the collision-less regime of the zero sound, i.e. $\om\gg \tau_{\rm col}^{-1}$. For the  equation of state with the isoterms, $P$ vs $1/n$,  having a van-der-Waals form, the incompressibility and the square of the first-sound velocity prove to be negative in the so-called spinodal region, $n_1 (T)<n<n_2(T)$. In case $P>0$ for all $n>0$, at a fixed averaged density (volume), $\bar{n}$, and temperature such that $n_1 (T)<\bar{n}<n_2 (T)$, after a while the system separates, as a result of the spinodal instability, in a part  in the gas phase with the density $n_{\rm G}$ and another part in the liquid phase with the density $n_{\rm L}$. The corresponding chemical potentials at constant temperature and pressure are equal in the equilibrium state, $\mu_{\rm G}=\mu_{\rm L}$,  forming the so-called Maxwell line in $P(1/n)$ dependence. The relative fraction of the gas phase $\chi$ ($0<\chi<1$) is determined by the averaged density $\bar{n} =n_{\rm G} \chi+n_{\rm L} (1-\chi)$. For a fixed averaged density $\bar{n}$ in the mixture, values $n_{\rm G}<n_1$, $n_{\rm L}>n_2$ and therefore both gas and liquid phases are stable as $f_{00} (n_{\rm G(L)})>-1$. With a decrease of $T$, isotherms $P(n)$ may cross the line $P(n)=0$ in two points for $n>0$, one corresponds then to a local maximum of the Landau free-energy and the other one to a local minimum. Collapsing to this minimum, the system becomes bound after a radiation of an energy excess, cf. \cite{SVB,Chomaz:2003dz}.

In the simplest case of ideal hydrodynamics the growing first-sound mode has the spectrum, cf. ~\cite{Skokov:2008zp-1,Skokov:2008zp-2,Randrup10-1,Randrup10-2,VS:2010gf-1,VS:2010gf-2}
\begin{align}\label{hydromode}
 -i\om =k\sqrt{|u_{S}^2| -ck^2}\,,
\end{align}
where $u_S$ is taken at fixed entropy, and $c$ is a coefficient related to the surface tension of droplets of one phase in the other one as $\sigma\propto \sqrt{c}$. For a non-zero thermal conductivity the isothermal $u_T$ and adiabatic $u_S$ first-sound velocities are different, and $u_T^2$ becomes negative at a higher $T$ than $u_S^2$, see Ref.~\cite{VS:2010gf-1,VS:2010gf-2}. After the replacement $u_S^2\to u_T^2$ Eq. (\ref{hydromode}) holds also for the case of a large thermal conductivity and small viscosity.
In the limit $T\to 0$ the isothermal and adiabatic first-sound velocities, $u_T$ and $u_S$, coincide.

Maximum of Eq.~(\ref{hydromode}) with respect to $k$ is
\begin{align}\label{spinod}
-i\om_m =v_{\rm F}p_{\rm F}(|f_{00}|-1)/(6m_{\rm F}\sqrt{c})\,.
\end{align}
The growth rate of the spinodal first-sound-like mode decreases with an increase of the surface tension of the droplets, whereas the growth rate of the collision-less zero-sound-like mode does not depend on the surface tension. For
\begin{align}\label{surftens}
\sqrt{c}>\frac{\pi|f_{00}|^{3/2}}{16 m_{\rm F} (|f_{00}|-1)^{1/2}}
\end{align}
the zero-sound-like excitations (\ref{omD-unst}), (\ref{maxsound}) would grow more rapidly than excitations of the  hydrodynamic mode (\ref{spinod}). Note that the presence of the viscosity  may delay formation of the hydrodynamic modes.

As we have mentioned in Introduction, at the first-order phase transition in  multi-component systems with charged constituents, like neutron stars, the resulting stationary state is a mixed pasta state, where finite size effects (a surface tension and a charge screening) are very important, cf. Refs.~\cite{Maruyama:2005vb,Tatsumi:2002dq-1,Tatsumi:2002dq-2}, contrary to the case of the one-component system, where the stationary state is determined by the Maxwell construction, cf. Ref.~\cite{SVB,Chomaz:2003dz}. For the isospin-symmetric nuclear matter at fixed temperature $T$, $T<15\mbox{--}20$ MeV, there exists a spinodal region in the dependence $P(1/n)$ at nucleon densities below the nuclear saturation density ($n_{1}(T)<n<n_2(T)<n_0$), see Ref.~\cite{SVB,Chomaz:2003dz}. The liquid-gas phase transition may occur in heavy-ion collisions. A nuclear fireball prepared in the course of a collision has a rather small size, typically less or of the order of the Debye screening length. Therefore, the pasta phase is not formed here.

\section{Bosonization of the local interaction}\label{sec:boson}

A fermionic system with a contact interaction can be equivalently described in terms of bosonic fields $\phi_{q}=\sum_{p} \psi^\dag_{p}\psi_{p+q}$, here we denote
$q=(\om,\vec{k})$ and $p=(\epsilon,\vec{p})$. In terms of the functional path integral the transition to a collective bosonic field can be performed with the help of the formal change of
variables by means of the Hubbard-Stratonovich transformation~\cite{Altland-Simons,Kopietz}. After this transformation the effective Euclidean interaction action for the system with
a repulsive interaction ($f_0 >0$) can be written within the
Matzubara technique in terms of the bosonic fields as
\begin{align}
S_{\rm int}[\phi]=\sum_{q} \frac{\phi_{q}\, \phi_{-q}}{2\Gamma_0^\om} - {\rm Tr}
\log \Big[1 -  T\hat{G}i\hat{\phi}_q\,\Big]\,,
\label{effact}
\end{align}
where $\hat{G}$ and  $\hat{\phi}$ are infinite matrices in
frequency/momentum space with matrix elements $[\hat
G]_{p,p'}=\delta_{p,p'} G(\epsilon,\vec{p}\,)$ and
$[\hat{\phi}]_{q,q'}=\delta_{q,q'} \phi_{\vec{q}}$. Trace is taken
over the Matsubara frequencies and momenta and includes factor 2 accounting for
the fermion spins. Transformation to zero temperature follows by
the standard replacement $T\sum_n \to\int d\epsilon/(2\pi
i)$.

Equation~(\ref{effact}) is applicable also for $f_0<0$ after the replacement $\phi\to i\phi$. Taking this into account and expanding Eq.~(\ref{effact}) up to the 4-th order in $\phi$ ($i\phi$) for
$T=0$ we obtain
\begin{align}
&S_{\rm int}[\phi] \approx\frac12 \mbox{sgn} (f_0) \sum_{q}
\phi_{q}\,\big((\Gamma_0^\om)^{-1} + a^2N\Phi(s,x)\big)\phi_{-q}
\label{effact4}\\ &\qquad +\frac14\sum_{\{q_i\}}U(q_1,q_2,q_3,q_4)
\phi_{q_1}\phi_{q_2}\phi_{q_3}\phi_{q_4}\delta_{q_1+q_2+q_3+q_4,0}
\,,\nonumber
\end{align}
where the effective field self--interaction is given by the
function
\begin{align}
U(q_1,q_2,q_3,q_4)
&=-2i\frac{1}{4!}\sum_{\mathcal{P}(1,\dots,4)}\sum_p
G_{p}G_{p+q_1} \nonumber\\ &\times
G_{p+q_1+q_2}G_{p+q_1+q_2+q_3}\,. \label{Lambda-gen}
\end{align}
Here the sum $\sum_{\mathcal{P}}$ runs over the 4! permutations
of four momenta $q_{i=1-4}$.

The general analysis of Eq.~(\ref{effact4})  for
arbitrary external momenta was undertaken in Ref.~\cite{Brovman-1,Brovman-2}.
The first term in Eq.~(\ref{effact4})  (quadratic in $\phi$) can
be interpreted as the inversed retarded propagator of the
effective boson zero-sound-like mode, cf. Eqs. (\ref{Tnr}), (\ref{TnrD}):
\begin{align}
\label{DS}
(D^R_{\phi})^{-1}(\om,k)= -{\rm sgn}(f_0)[(\Gamma_0^\om)^{-1} +
a^2N\Phi(s,x)]\,.
\end{align}

To describe modes for an attractive interaction, $f_0<0$, it is
instructive to re-derive the expression for the mode propagator
using the approach proposed in Ref.~\cite{IKV00}. The local
four-fermion interaction $\Gamma_0^\om$  can be viewed as an
interaction induced by the exchange of  a scalar heavy boson with
the mass $m_{\rm B}$,
\begin{align}
\label{Gamom} \Gamma_0^\om \to -\frac{\Gamma_0^\om m_{\rm
B}^2}{(\om+i0)^2-k^2-m_B^2}\equiv -\Gamma_0^\om\, m_{\rm
B}^2\,D^R_{\rm B,0}(\om,k)\,,
\end{align}
where $D^R_{\rm B,0}$ is the bare retarded Green function of the
 heavy boson and we assumed that $m_{\rm B}^2$ is much larger
then typical squared frequencies and momenta, $\om^2, k^2$, in the
problem.

Then making the replacement (\ref{Gamom}) in  the fermion
scattering amplitude (\ref{Tph-sol}) we obtain
\begin{align}
T^R_{{\rm ph},0}(\om,k)&= \frac{-(a^2N)^{-1}f_0 m_{\rm
B}^2}{(D_{{\rm B}, 0}^R)^{-1}(\om,k) - \Sigma^R_{\rm B}(\om,k)  }
 \equiv  V_{\rm B}^2 D^{\rm R}_{\rm B}(\om,k)\,.
\label{Tph-sol1}
\end{align}
Hence, we can identify the vertex of the boson-fermion interaction
\begin{align}\label{vert}
V_{\rm B}^2= - f_0 m_{\rm B}^2/(a^2N)\,,
\end{align}
and the full retarded Green function of the boson $D^R_{\rm B}(\om,k)$ with the retarded self-energy
\begin{align}
\label{SigmaR}
\Sigma^R_{\rm B}(\om,k) = f_0 m_{\rm B}^2 \Phi\big(\frac{\om}{kv_\rmF},\frac{k}{p_\rmF}\big)\,.
\end{align}
Note that for very large $m_{\rm B}$ that we have assumed, $D^{\rm
R}_{\rm B}(\om,k)$ in (\ref{Tph-sol1}) differs from $D^{\rm
R}_{\phi}(\om,k)$ in (\ref{DS}) only by a frequency-independent prefactor. For an
attractive interaction, which we are now interested in ($f_{0}<0$),
we have $m_{\rm B}^2 >0$ and $V_{\rm B}^2>0$. For a repulsive interaction instead of the bare scalar boson the bare vector boson would be an appropriate choice, cf. ~\cite{IKV00}, or we may come back to the formalism given by Eqs. (\ref{effact})--(\ref{Lambda-gen}).

The spectrum of the zero-sound-like excitations follows from the solution of the Dyson equation $0=[D_{\rm
B}^R(\om,k)]^{-1}\approx -m_{\rm B}^2(1+f_{0}\Phi(s,x))$ for
$m_{\rm B}\gg \om,k$ and coincides with the solution of
Eq.~(\ref{sound-eq}).
The boson spectral function is given by
\begin{align}
&A_{\rm B}(\om,k)=-2\Im D_{\rm B}^R(\om,k), \nonumber\\ &-2\Im
T^R_{{\rm ph},0}(\om,k)=V_{\rm B}^2 A_{\rm B}(\om,k)\,.
\end{align}

\section{Bose condensation in the mean-field approximation}\label{sec:Bosecond}

\subsection{Condensation of a scalar field $\phi$ for $f_0<-1.$}

Consider a Fermi liquid with a local scalar interaction $f_0<-1$. We assume that the bosonization procedure of the interaction, Eqs.~(\ref{Gamom}) and~(\ref{Tph-sol1}), is performed. According
to the perturbative analysis in Sect.~\ref{sec:pominstab}, for $f_0<-1$ there are modes which grow with time. The growth of hydrodynamic modes (first sound), cf. Eqs.~(\ref{hydromode}) and (\ref{spinod}), results in the formation of a mixed phase. Besides the hydrodynamic modes the zero-sound-like modes grow with time, cf. Eqs.~(\ref{omD-unst}) and (\ref{maxsound}). We study now the opportunity that the instability of the zero-sound-like modes at densities where $f_0<-1$ may result in the formation of a condensate of the scalar field $\phi$. The latter quantity is to be understood as the expectation value, $\phi=\langle\hat{\phi}\rangle$, of the field operator $\hat{\phi}$ of the scalar quanta of particle-hole excitations determined by the local interaction described in the Fermi-liquid approximation by the Landau parameter $f_0$. This may lead to a decrease of the system energy and to a rearrangement of the excitation spectrum on top of the condensate field.

We will exploit the simplest trial function describing the complex scalar field of the form of a running wave
\begin{align}\label{running}
\phi =\phi_0 e^{-i\om_c  t+i\vec{k}_c\vec{r}}\,,
\end{align}
with the condensate frequency and momentum $(\om_c, \vec{k}_c)$ and the constant amplitude $\phi_0$. The choice of the structure of the order parameter is unimportant for our  study of the stability of the system.

Guided by the construction of the full Green function of the
effective boson field, see Eq.~(\ref{Tph-sol1}), the effective
Lagrangian density for the condensed field (\ref{running}) can be
written  as, cf. Ref.~\cite{MSTV90,V84-1,V84-2},
\begin{align}
\label{Lagr-B} L=\Re D_{\rm B}^{-1} (\om_c, k_c) |\phi_0|^2
-{\textstyle\frac12}V_{\rm B}^4\Lambda(\om_c, k_c)
|\phi_0|^4\,.
\end{align}
Here we assume that the mean-field value $|\phi_0|^2$ is rather small and restrict the expansion by the $|\phi_0|^4$ term.

The energy density of the condensate is given then by the standard
relation
\begin{align}
E_{\rm B} =\om \partial L/\partial \om -L \,.
\nonumber
\end{align}
Fully equivalently, this Lagrangian density can be written in the
form suggested by expansion~(\ref{effact4}), now applied for
the running wave classical field. After the field redefinition
$\phi_0\to (a/m_{\rm B})\sqrt{N/|f_0|}\phi_0=\phi_0/|V_{\rm B}|$,
\begin{align}
\label{Lagr} L = \Re D_{\phi}^{-1}(\om_c,k_c) |\phi_0|^2
 -{\textstyle\frac12}\Lambda(\om_c, k_c) |\phi_0|^4\,.
\end{align}
Here and in Eq.~(\ref{Lagr-B}) the quantity $\Lambda(\om_c, k_c)$
represents the self-interaction amplitude of condensed modes which
corresponds to the ring diagram with four fermion Green
functions (\ref{Lambda-gen}),
$\Lambda(\om_c,k_c)=6U(q_c,-q_c,q_c,-q_c)$.  As we shall see below
the energetically favorable is the state with $\om_c =0$, and
therefore we put $\om_c=0$ here. The leading order contribution to
the self-interaction parameter $\Lambda(0,k_c)$ as a function of
the condensate momentum was calculated in Ref.~\cite{Brovman-1,Brovman-2}:
\begin{align}
\Lambda(0,k_c)&=\frac{8 a^4{\nu}}{\pi^2 v_\rmF^3 x_c^4}\Big[
\frac{1-x_c^2}{x_c}\ln\Big|\frac{2+x_c}{2-x_c}\Big| -
\frac{1-x_c^2}{2x_c}\ln\Big|\frac{1+x_c}{1-x_c}\Big| \nonumber\\
&+ \frac{x_c^2}{4-x_c^2}\Big]\,,\quad x_c=k_c/p_\rmF.
\label{Lambda}
\end{align}
For $x_c\ll 1$ we get
\begin{align}
\Lambda(0,k_c)\approx
a^4\lambda\,\Big(1+\frac{k_c^2}{2p_\rmF^2}\Big)\,,\quad \lambda =
\frac{{\nu}}{\pi^2v_\rmF^3}\,. \label{lamb-c}
\end{align}
The quantity $\lambda$ agrees also with the result derived in
Ref.~\cite{D82-1,D82-2} for description of the pion condensation in the
Thomas-Fermi approximation.

The equation of motion for the static field amplitude follows from the
variation of the action corresponding to the Lagrangian density
(\ref{Lagr}),
\begin{align} \label{eq4fi0}
-a^2N\tilde{\om}^2(k_c)\phi_0 -\Lambda(0,k_c)\,|\phi_0|^2\phi_0 =
0\,,
\end{align}
where we introduce the effective boson gap
\begin{align}\label{omti}
\tilde{\om}^2(k_c)=[1-|f_0(k_c)|\Re\Phi(0,k_c)]/|f_0(k_c)|\,,
\end{align}
as it was done in the description of the pion condensation, see Ref.~\cite{Migdal78,MSTV90}.

The equation for the static condensate amplitude (\ref{eq4fi0}) has a non-trivial solution for
$\tilde{\om}^2(k_c)<0$,
\begin{align}
|\phi_0|^2 = \phi_{0,\rm eq}^2 =-N\frac{\tilde{\om}^2(k_c)}{
 a^2\lambda\,\Big(1+\frac{k_c^2}{2p_\rmF^2}\Big)}\,\theta(-\tilde{\om}^2(k_c)) \,,
\label{varphi0}
\end{align}
which corresponds to the gain in the energy density
\begin{align}\label{Eb}
E_{\rm B}=- N^2\frac{\tilde{\om}^4(k_c)}{ 2
\lambda\,\Big(1+\frac{k_c^2}{2p_\rmF^2}\Big)}\theta(-\tilde{\om}^2(k_c))\,,
\end{align}
where $\theta (x)=1$ for $x>0$ and 0 otherwise.

To describe a slow dynamics of the condensate  one may generalize Eq.~(\ref{eq4fi0}) for  $\hat{\omega}=i\partial_t$ and $\hat{k}=k_c-i\nabla$~\cite{MSTV90,Ivanov:2000ma}:
\begin{align}\label{eq4fi01}
D_{\phi}^{-1}(\hat{\omega}, \hat{k})\phi_0 -\Lambda(0,k_c)\,|\phi_0|^2\phi_0 =
0\,,
\end{align}
where the  condensate amplitude is now assumed to be a smooth  function of $t$ and $\vec{r}$.

Thus, in the perturbative analysis (for a small $|\phi_0|^2$) we have demonstrated that the appearance of the condensate is energetically profitable at least provided the density is fixed. In a more realistic situation the volume is fixed that corresponds to a fixed averaged density rather than to a local one. In this case the  development of the condensate amplitude occurs simultaneously with the process of the phase separation and the question about the structure of the resulting equilibrium state is more cumbersome and needs a further investigation.

In the presence of the weak condensate there appears an additional contribution
$f_0^{\rm B}(k_c)$ to the scalar Landau parameter $f_0(k)$. It can be expressed through the incompressibility
\begin{align}
K=K_{\rm F}+K_{\rm B}=n\frac{d^2 E_{\rm F}}{d n^2}+n\frac{d^2 E_{\rm B}}{d n^2}
\,.
\label{K1}
\end{align}
On the other hand now
\begin{align}
\delta \mu_{\rm F} =\frac{2\epsilon_{{\rmF}}}{3n} \delta n
+ \frac{f_{0}+f_{0}^{\rm B}}{N}\delta n \,,
\end{align}
and therefore
\begin{align}
K=\frac{p^2_{\rm F}}{3m_{\rm F}^*}\big(1+f_{0}+f_{0}^{\rm B}\big)\,. \label{Kcond}
\end{align}
Here $f_0$ and $f_0^{\rm B}$ are functions of $k_c$ with $k_c$ to be found from the energy minimization. So, if one knows $E_{\rm F}(n)$, using Eqs.~(\ref{Eb}) and (\ref{K1}), (\ref{Kcond}) one may find $f_0$, $f_0^{\rm B}$ in the presence of the condensate:
\begin{align}
f_{0}= \frac{3 n}{2\epsilon_{\rm F}}\frac{d^2 E_{\rm F}}{d n^2}-1,\quad
f_{0}^{\rm B}= \frac{3 n}{2\epsilon_{\rm F}}\frac{d^2 E_{\rm B}[f_{0}^{\rm tot}]}{d n^2}\,,
\label{f0-fB}
\end{align}
where $f_{0}^{\rm tot}=f_{0}+f_{0}^{\rm B}$ obeys the differential equation
\begin{align}
f_{0}^{\rm tot}=f_{0}+\frac{3 n}{2\epsilon_{\rm F}}\frac{d^2 E_{\rm B}[f_{0}^{\rm tot}]}{d n^2}\,.
\label{diff-ftot}
\end{align}
In  case of a weak condensate (for $|f_{0} +1|\ll 1$) one can use $E_{\rm B}[f_{0}^{\rm tot}]\approx E_{\rm B}[f_{0}]$. For a developed condensate the  perturbative analysis does not work and one should take into account that  $E_{\rm B}=E_{\rm B}[f_0^{\rm tot}]$ and solve Eq.~(\ref{diff-ftot}) self-consistently.

Recall that in the mean-field approximation, i.e. neglecting the reconstruction of the sound spectra on top of the new condensate, the medium remains unstable with respect to the growth of the zero-sound  and of the first-sound modes, since $f_0^{\rm tot} <-1$.

\subsection{Bose condensation in a homogeneous state.}

If the minimum of the gap $\tilde{\om}^2$ is realized at $k_c=0$, e.g., it is so for $f_{02}\ge 0$, the energy density $E_{\rm B}$ is minimized for $\om_c=k_c=0$. Then, the equilibrium condensate amplitude and energy density are
\begin{align}
\label{varphi0-00}
\phi_{0,\rm eq}^2 = N\frac{z_f}{a^2
\lambda}\,\theta(z_f)\,, \quad E_{\rm B}=-N^2\frac{z_f^2}{2
\lambda}\,\theta(z_f) \,.
\end{align}
The Bose condensate is formed for $f_{00}<-1$. For $|f_{00} +1|\ll 1$ the value $z_f (f_{00})$ is determined by Eq.~(\ref{sd0}). For the developed condensate $z_f =z_f (f_{00}^{\rm tot})=1-1/|f_{00}^{\rm tot}|$.

As follows from Eqs.~(\ref{PhiLow}) and (\ref{eq4fi01}), the amplitude of the homogeneous condensate field starts growing according to Eq.~(\ref{omD-unst}) with  $k_c\sim 1/\xi$, where $\xi $ is the condensate coherence length. The latter can be estimated by setting
\begin{align}
D^{-1}_{\phi}(0, \hat{k})=a^2 N \Big[|f_{00}|^{-1}-1 -\frac{1}{12 p_{\rm F}^2}\nabla^2\Big]
\end{align}
in Eq.~(\ref{eq4fi01}). Thus, we get $\xi \sim \sqrt{\frac{|f_{00}|}{12p_{\rm F}^2(|f_{00}|-1)} }$ (valid only if $|f_{00}|-1\ll 1$).
Comparing Eqs.~(\ref{omD-unst}) and (\ref{spinod}), we find that the condensate grows faster than aerosol structures in the spinodal region if
\begin{align}\label{hk}
\sqrt{c}\gsim  \frac{|f_{00}|^{3/2} }{13 m_{\rm F}(|f_{00}|-1)^{1/2}}\,.
\end{align}

For the case of non-zero temperature, $T\neq 0$, in the mean-field
approximation one should just replace $\Phi\to \Phi_T$, where the
fermionic step-function distributions are replaced by the thermal
distributions. For $T\ll \epsilon_\rmF$, from Eq.~(\ref{PhiT}) one then recovers the equation for the critical temperature of the condensation:
$$T^{\rm MF}_c/\epsilon_\rmF=\sqrt{12\,z_f (f_{00}^{\rm tot}(T^{\rm MF}_c))}/\pi$$ valid for $0<z_f\ll
1$. For $T\gg \epsilon_{\rm
F}$ and $|f_{00}^{\rm tot}|\gg 1$ from Eq.~(\ref{PhiT1}) one would find $T^{\rm
MF}/\epsilon_{\rm F} = 2|f_{00}^{\rm tot}(T^{\rm MF}_c)|/3$.

The change of the free energy density owing to the condensation for $T\neq 0$ is, cf. Eq.~(\ref{varphi0-00}),
\begin{align}
F_{\rm B}(n,T)=-3\nu\,n \,\epsilon_{\rm F}\frac{(|f_{00}^{\rm tot} (n,T)|-1)^2}{(f^{\rm tot}_{00}(n,T))^2}\,\theta(-1-f_{00}^{\rm tot}) \,.
\label{FB}
\end{align}

Recall that $u_T^2$ is
negative within the spinodal region which exists, if the pressure
isotherm has a van-der-Waals form. Thus, the formation of the scalar Bose
condensate, suggested here, might compete with the development of
the spinodal instability at the liquid-gas phase transition in
Fermi liquids. At least, if the surface tension between liquid and gas regions is
sufficiently high, the aerosol-like mixture appearing in the course of the development of the spinodal decomposition evolves more slowly than the scalar condensate field. If the free-energy gain owing to the condensation were rather pronounced there could appear a local minimum in the Landau free energy corresponding to a metastable state at a density $n_{\rm ms}$, $n_1<n_{\rm ms}<n_2$.

\subsection{Bose condensation in an inhomogeneous state.}

We discuss now how the momentum dependence of the coupling constant $f_0(k)$, as given by Eq.~(\ref{fexp}), may influence the condensation of the scalar mode. The  conditions for the appearance of a condensate with $\om_c=0$ and a finite momentum $k_c$ are determined by relations $\tilde{\om}^2(k_c)\le 0\,,\quad \frac{\rmd}{\rmd k}\tilde{\om}^2(k_c)=0$\,. The critical condition for the appearance of the condensate can be deduced from the expansion of $|f_0 (k)|\tilde{\om}^2(k)$ in Eq.~(\ref{omti}) in $k^2$ for $x^2=k^2/p_{\rm F}^2\sim x_m^2$:
\begin{align}\label{crpar}
&1+f_0(x)\Phi(0,x)\approx 1 +f_{00} + 3\frac{
(f_{02}-\frac{f_{00}}{12})^2}{(f_{02} + \frac{f_{00}}{20})}
\nonumber\\
&-\frac{ f_{02} +\frac{f_{00}}{20}  }{12} [x^2 - x_m^2]^2\,,
\quad
x_m^2=6 \frac{f_{02}-\frac{f_{00}}{12}}{f_{02} + \frac{f_{00}}{20}}.
\end{align}
For $f_{00}<-1$ this function has always a negative minimum at
$x=0$, the second and deeper minimum appears at finite $x$, if
\begin{align}
f_{02}\le f_{02}^{\rm cr, nec}={f_{00}}/{12}\,. \label{cond1}
\end{align}
Thus, we derived the necessary condition for the appearance of an inhomogeneous condensate at $f_{00}<-1$. The sufficient condition can be derived from the requirement that the energy gain owing to the inhomogeneous condensate given by Eq.~(\ref{Eb}), is larger than that for the homogeneous condensate given by Eq.~(\ref{varphi0-00}). We find that the inhomogeneous condensation is more energetically favorable compared to the homogeneous one if
\begin{align}
f_{02}\le f_{02}^{\rm cr}=-{f_{00}^2}/{3}-{f_{00}}/{4}\,.
\label{cond3}
\end{align}

\begin{figure}
\centerline{\includegraphics[width=7cm]{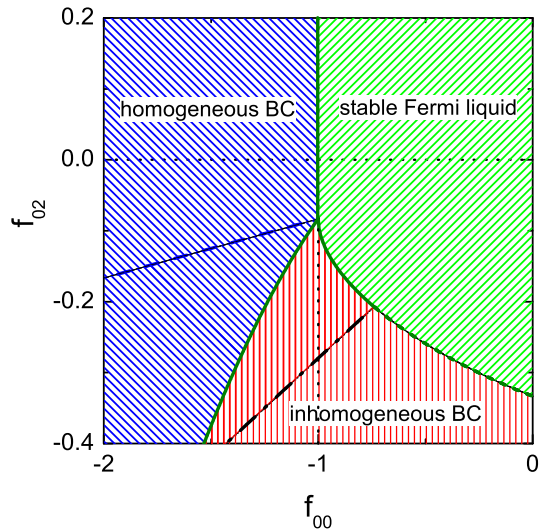} }
\caption{(Color online) Phase diagram of a Fermi liquid with a scalar interaction
in the particle hole channel described by the momentum dependent
coupling constant $f_0(k/p_\rmF)=f_{00}+f_{02}k^2/p_\rmF^2$.
Depending on the values of parameters $f_{00}$ and $f_{02}$ the
Fermi liquid can be either in a stable state or in  states with
homogeneous or inhomogeneous static Bose condensates (BC) of
scalar modes.
Left solid line shows the critical value of $f_{02}$ given by Eq.~(\ref{cond3}).
Right solid line and its short-dashed continuation depict the critical value of $f_{02}$
given by Eq.~(\ref{cond2}).
Vertical solid line represents the border $f_{00}=-1$.
Long-dashed line shows the necessary condition for the inhomogeneous  condensation (\ref{cond1}).
Dash-dotted line is the solution of equation $f_{02}x_m^2 =f_{00}$, where $x_m$ is given by
Eq.~(\ref{crpar}).
A possibility of the development of a spinodal instability is disregarded here.} \label{fig:PhaseD}
\end{figure}

For $-1<f_{00}<0$ the minimum at finite momentum can be realized if
\begin{align}
f_{02}\le  f_{02}^{\rm cr}=-\frac16\Big[1+\half
f_{00}+\sqrt{1+\frac{2}{5}f_{00}-\frac35 f_{00}^2}\Big]\,,
\label{cond2}
\end{align}
and
\begin{align}
x_m^2(f_{02}^{\rm cr})=6\frac{1+f_{00}+\sqrt{1+\frac{2}{5}f_{00}-\frac35
f_{00}^2}}{1+\sqrt{1+\frac{2}{5}f_{00}-\frac35 f_{00}^2}+\frac15
f_{00}}\,.
\end{align}
The solid lines in Fig.~\ref{fig:PhaseD} on the
$(f_{00},f_{02})$ plane show the  critical values  $f_{02}^{\rm
cr}$ given by Eqs. (\ref{cond3}) and (\ref{cond2}). Together with the line $f_{00}=-1$ they form the phase diagram of the Fermi liquid with the momentum dependent
interaction in the scalar channel. Three phases are shown here: stable Fermi liquid, Fermi liquid with a homogeneous Bose condensate of zero-sound-like modes, and Fermi liquid with an inhomogeneous Bose condensate. A  possible spinodal instability is not considered here. We see that for $f_{02}>-1/12$ there are only two phases possible:
stable Fermi liquid and homogeneous Bose condensate. The latter is formed if $f_{00}<-1$.
For $f_{02}>-1/12$ all three phases are possible depending on the value of $f_{00}$. With a decrease of $f_{00}$ the Fermi liquid develops first an inhomogeneous condensate, which converts then in a homogenous condensate with a further $f_{00}$ decrease.
Auxiliary lines shown in Fig.~\ref{fig:PhaseD} are the necessary condition for an inhomogeneous condensation given by Eq.~(\ref{cond1}) (long-dashed line) and the solution of equation $f_{02}x_m^2 =f_{00}$ (dash-dotted line). In the region restricted by this dash-dotted line and the short-dashed continuation of the solid line we have $|f_{00}|< |f_{02}|x_m^2 $ and the
expansion (\ref{fexp}) for $f_0 (k)$ becomes questionable.

Near the minimum the boson gap  $\tilde{\om}^2(k)$ can be presented as
\begin{align}
\tilde{\om}^2 (k) =\tilde{\om}^2_0 (k_c) +\frac{{\gamma}}{4k_c^2} (k^2 -k_c^2)^2\,,
\label{gap-min-exp}
\end{align}
for $\gamma ={\rm const}$. The critical point is found from the
condition  $\tilde{\om}^2_0 (k_c=k_{\rm cr})=0$. Near the critical point the parameters in Eq.~(\ref{gap-min-exp}) follow from Eq.~(\ref{crpar}). Beyond the critical point the actual value of $k_c$ follows from the minimum
of the energy density (\ref{Eb}) as a function of $k_c$.

As follows from Eqs.~(\ref{PhiLow}) and (\ref{eq4fi01}), the beginning of the growth of the condensate field amplitude follows the law
\begin{align}
\omega =i 2k_c v_{\rm F}\tilde{\om}^2_0 (k_c)/\pi \,,
\end{align}
for $\tilde{\om}^2_0 (k_c)<0$.
Comparing this frequency with Eq.~(\ref{spinod}) we find that the inhomogeneous condensate field will grow more rapidly compared to the growth of the aerosol droplets in spinodal region provided the surface tension fulfills inequality
\begin{align}\label{homogk}
\sqrt{c}>\frac{\pi p_{\rm F}\left[|f_0 (k_m =|u_T|/\sqrt{2c})|-1\right]}{12 m_{\rm F} k_c |\tilde{\om}^2_0 (k_c)|}\,.
\end{align}

\section{Avoiding the Pomeranchuk instability}\label{sec:pomer}

\subsection{Reconstruction of the excitation spectrum on top of the condensate}\label{sec:Cond-rec}

Let us now study excitations in the presence of the condensate. Setting $\phi_q =\phi +\phi'_q$ with $\phi$ as a mean field in the interaction action (\ref{effact4}) and varying the latter with respect to $\phi'_q$ keeping only terms linear in $\phi'_q $ we recover the spectrum of over-condensate zero-sound-like excitations.
\begin{align}
a^2 N\Big[-|f_{0}^{\rm tot}(k)|^{-1} + \Phi\Big(\frac{\om}{k v_\rmF}, \frac{k}{p_\rmF}\Big)\Big] - \delta\Sigma_\phi=0,
\label{dispeq-cond}\end{align}
where the last term
\begin{align}
\delta\Sigma_\phi (\omega , k, \phi)&= 2\Lambda(\om,k)\,|\phi|^2
\label{dSigfi1}
\end{align}
arises from the interaction of excitations with the condensate. The same equation is obtained  after replacement $\phi\to \phi +\phi^{'}$ in Eq.~(\ref{eq4fi01}) and its linearization.
Making use of the equilibrium value of the condensate field from Eq.~(\ref{varphi0-00}), we obtain
\begin{align}
\delta\Sigma_\phi (\omega, k, \phi_{0,\rm eq}) = -2\,a^2N\,\tilde{\om}^2(k_c)\Lambda(\omega,k)/\Lambda (0,k_c).
\label{dSigfi}
\end{align}
This result holds both for $k_c \neq 0$ and $k_c =0$.
Using expansion Eq.~(\ref{PhiLow}) we find the spectrum for $0<|\tilde{\om}^2(k_c)|\ll 1$,
\begin{align}
\label{newspectr}
{\om} \approx i\frac{2}{\pi} \tilde{\om}^2(k_c) \,k v_{\rm F}\,\,\,{\rm for}\,\,\,
\tilde{\om}^2(k_c)<0\,,
\end{align}
and observe that the overcondensate zero-sound-like excitations are damped, similar to those we obtain for the case $-1<f_{00}<0$, see Eq.~(\ref{omnegf}). \emph{Thus, we demonstrated that in the presence of  the scalar condensate field $\phi_{\rm 0, eq}$ the Fermi liquid is
free from the Pomeranchuk instability of the zero-sound-like modes.}

The particle-hole interaction also changes in the presence of the condensate.
There appears a new term in Eq.~(\ref{Tph-sol}) for the particle-hole amplitude, which now reads as
\begin{align}
&a^2\,N\,T_{\rm ph,0}^R (\omega, k,\phi)
\nonumber\\
&\qquad =\Big[\frac{1}{f_0^{\rm tot}(k)}+
\Phi (\omega,k)-\frac{\delta\Sigma_\phi(\omega,k,\phi)}{a^2N} \Big]^{-1}
\nonumber \\
&\qquad =
\Big[\frac{1}{f_{\rm ren, 0}(\omega, k,\phi)}+\Phi (\omega,k) \Big]^{-1}\,.
\label{Tph-new}
\end{align}
In the last relation we introduced the renormalized local interaction
\begin{align}
f_{\rm ren,0}(\omega, k,\phi) =
\Big[\big[f_{0}^{\rm tot}(k)\big]^{-1} - 2\Lambda (\omega,k)\frac{|\phi^2|}{a^2N}\Big]^{-1}
\,,
\nonumber
\end{align}
and in our scheme $f_{0}^{\rm tot}(k_c)$ is determined by Eq.~(\ref{diff-ftot}).
Substituting the equilibrium values of the condensate momentum $k_c$ and amplitude $\phi_{0,\rm eq}$, which follow from the energy minimization, we find
\begin{align}
&f_{\rm ren,0}= f_{\rm ren,0}(0, k_c,\phi_{0,\rm eq})=
 \frac{f_{0}^{\rm tot}(k_c)}{1 + 2 f_{0}^{\rm tot}(k_c)\, \tilde{\om}^2(k_c)}\,.
\label{fren1}
\end{align}
Knowing the value $f_{\rm ren,0}$ one can reconstruct the energy density ${E}_{\rm tot}(n)$ of the Fermi liquid from the differential equation
\begin{align}
\label{frenE}
\frac{3n}{2\epsilon_{\rmF}}\frac{\rmd^2 {E}_{\rm tot}(n)}{\rmd n^2}-1=f_{\rm ren,0}\,.
\end{align}
This energy density includes both mean-field and quadratic-fluctuation contributions.

As we have demonstrated, after accumulation of the equilibrium condensate field, the Fermi liquid becomes stable with respect to the zero-sound-like excitations. Now from Eq.~(\ref{fren1}) we see that for $f_{0}^{\rm tot}(k_c)<-1 - k_c^2/(6p_{\rm F}^2)+O(k_c^4/p_{\rm F}^4)$ one has
$f_{\rm ren,0}>-1$. If so, the appearance of the condensate makes the first-sound modes stable, as their speed squared is
\begin{align}
u^2 =\frac{p_{\rm F}^2}{3m_{\rm F} m^{*}_{\rm F}}
(1+f_{\rm ren,0})>0\,.
\end{align}
However, for $k_c\neq 0$ there might exist a narrow interval, $-1 - k_c^2/(6p_{\rm F}^2)< f_{0}^{\rm tot}(k_c)< f_0^{\rm tot, cr}<-1$, where the zero-sound is stable, but the first sound is  unstable. Here  $f_0^{\rm tot, cr}$ is calculated following Eq. (\ref{fren1}) with $f_0^{\rm cr}=f_{00}+f_{02}^{\rm cr} x_{m}^2(f_{00}, f_{02}^{\rm cr})$, with $x_m^2(f_{00}, f_{02}^{\rm cr})$ from (\ref{crpar}) and $f_{02}^{\rm cr}$ determined by   condition (\ref{cond3}). To answer the question whether the mentioned interval exists one needs a self-consistent solution of the problem. Since information on the quantity  $f_{02}$ is poor,  below we focus on a homogeneous condensation.

In the case of a homogenous condensate with $k_c=0$ we have $\tilde{\om}^2(k_c)=-z_f=
-1-1/f_{00}^{\rm tot}$ and Eq.~(\ref{fren1}) reduces to
\begin{align}
f_{\rm ren,00} = -\frac{f_{00}^{\rm tot}}{2f_{00}^{\rm tot}+1}\,.
\label{fren}
\end{align}
As seen from Eqs.~(\ref{fren1}) and (\ref{fren}), if originally $f_{00}<f_{00}^{\rm cr}=-1$ and therefore $f_{00}^{\rm tot}<-1$, the renormalized interaction $-1<f_{\rm ren,00} < -1/2$ and the first sound is stable.

\emph{Thus, we demonstrated that  for $f_{00}<-1$ in case  of  the homogeneous scalar condensate field, the Fermi liquid is free from the Pomeranchuk instability of the first-sound-like modes.}

Several approximations are beyond our considerations above.
In the action (\ref{effact4}) and the Lagrangian density (\ref{Lagr-B}) and (\ref{Lagr}) we restricted the expansion by $\phi^4$ terms. We disregarded interactions of excitations on top of the condensate, neglecting $\phi'^{3}$, $\phi'^{4}$ terms in the Lagrangian. Moreover, exploiting (\ref{varphi0}) we disregarded in Eq.~(\ref{dSigfi}) feedback of fluctuations on the mean field $\phi$. Thus, the value $f_{\rm ren,0}$  given by Eq.~(\ref{fren1}), which is found with the help of $f_{\rm tot,0}$ determined in Eq.~(\ref{diff-ftot}), is consistent with that follows from Eq.~(\ref{frenE}) only if on the one hand the $\phi_0$ mean field is rather small, i.e. for $|f_{0}-f_0^{\rm cr}| \ll 1$
and on the other hand the fluctuation terms on the top of the condensate yield a yet smaller contribution. Our procedure here is similar to that exploited in the standard Landau's phase transition theory, cf.~\cite{LP1981,Tilly-Tilly}, where one expands a thermodynamical potential in the order parameter up to $\phi^4$ and treats fluctuations of the order parameter perturbatively.  The approximation is valid for ${\rm Gi}=|T_{\rm fl}-T_c|/T_c\ll |T-T_c|/T_c\ll 1$, where $T_c$ is the critical temperature of the second-order phase transition and ${\rm Gi}$ is the Ginzburg number. The temperature $T_{\rm fl}$ determines the so-called fluctuation region near the critical point.

In the general case $f_{\rm ren,0}$ should include  infinite series of more complicated diagrams. Then the condensate field  follows from the minimization of full energy density ${E}_{\rm tot}(n)$ including also fluctuations. The full $f_{\rm ren,0}$ would have a complicated density dependence related to the full energy density of the Fermi liquid following Eq.~(\ref{frenE}).

Additionally, our estimation of the condensate self-interaction parameter $\Lambda$ in Eqs.~(\ref{Lambda}) and (\ref{lamb-c}) was done in the leading one-loop order only. Inclusion of other diagrams could change the value $\Lambda(0,k_c)$ which determines the condensate amplitude in Eq.~(\ref{eq4fi0}). We notice that if $\Lambda(0,k_c)$ occasionally became $\gg 1$ then following Eq.~(\ref{diff-ftot}) would be $f_{0}^{\rm tot}\simeq f_{0}(n)$.
Formally then in case of a homogeneous scalar condensate the renormalized Landau parameter would take the value $f_{\rm ren,0}\to f_{\rm ren,0}^{(\infty)}=-f_{0}(n)/(1+2\,f_{0}(n))$
independently on parameters related to the  condensate, being thus fully determined by fluctuations on top of the condensate. In reality such a limit case is not realized, since when fluctuations yield contribution to the energy of the same order as the mean field, the condensate amplitude and the over-condensate excitations should be treated self-consistently as in the case of the dynamical spontaneous symmetry breaking considered in Ref.~\cite{Coleman-Weinberg}.

\subsection{A possibility of a new metastable state in isospin-symmetric nuclear matter for $n<n_0$}

We consider now an example of a two-component Fermi liquid  -- the isospin-symmetric nuclear matter for $n<n_0$ and $T=0$. A simple phenomenological parametrization for the volume part of the energy per baryon  was proposed in Ref.~\cite{KFW02} as an expansion in powers of $p_{\rm F}/m_N$
\begin{eqnarray}
\mathcal{E}_{N,\rm v}(n)=\frac{3p_\rmF^2}{10\,m_N} - c_1\frac{p_\rmF^3}{m_N^2}
+ c_2 \frac{p_\rmF^4}{m_N^3} + c_3 \frac{p_\rmF^5}{m_N^4}
\,,
\label{weise-exp-2}
\end{eqnarray}
where $m_N=939$\,MeV is the free nucleon mass and parameters $c_1$, $c_2$ and $c_3$ are expressed through values of the binding energy $-\mathcal{E}_{N,\rm v}(n_0)$, incompressibility $K_{\rm v}(n_0)$ of the nuclear matter and the nuclear saturation density $n_0$,
\begin{align}
c_1\frac{p_{\rmF}^3(n_0)}{m_N^2} &= - 10 \mathcal{E}_{N,\rm v} (n_0) -\frac92 K_{\rm v}(n_0)
+ \frac{9}{10} \frac{p^2_\rmF(n_0)}{m_N}\, ,
\nonumber\\
c_2\frac{p_{\rmF}^4(n_0)}{m_N^3} &= -15 \mathcal{E}_{N,\rm v}(n_0) - 9 K_{\rm v}(n_0) + \frac{9}{10}\frac{p^2_\rmF(n_0)}{m_N} \, ,
\nonumber\\
c_3\frac{p_{\rmF}^5(n_0)}{m_N^4} &= \phantom{-} 6\, \mathcal{E}_{N,\rm v}(n_0) +\frac92 K_{\rm v}(n_0) -
\frac{3}{10} \frac{p^2_\rmF(n_0)}{m_N}\,,
\nonumber
\end{align}
$K_{\rm v}(n)=n\rmd^2(n\mathcal{E}_{N,\rm v})/\rmd n^2$.
Taking appropriate values $n_0=0.17\,{\rm fm}^{-3}$, $\mathcal{E}_{N,\rm v}(n_0)=-16$\,MeV and
$9K_{\rm v}(n_0)=285$\,MeV we find
\begin{align}
c_1 &= 3.946\,,
\quad
c_2 = 3.837\,,
\quad
c_3 = 13.10\,.
\nonumber
\end{align}
From Eq.~(\ref{weise-exp-2}) the nuclear incompressibility and the corresponding $f_{00}$
parameter follow as
\begin{align}
&K_{\rm v}(n) = \frac{p_\rmF^2}{3 m_N}
- c_1 \frac{2 p_\rmF^3}{m_N^2}
+ c_2 \frac{28 p_\rmF^4}{9 m_N^3}
+ c_3 \frac{40 p_\rmF^5}{9 m_N^4}\,,
\label{K_N}\\
&f_{00,\rm v}(n)  =
\Big(\frac{m_N^*}{m_N}-1\Big)
\nonumber\\
&\quad + \frac{m_N^*}{m_N}\frac{p_\rmF}{m_N}
\Big( -6 c_1
+c_2 \frac{28 p_\rmF}{3 m_N}
+c_3 \frac{40 p_\rmF^2}{3 m_N^2}
\Big)\,.
\label{f0_N}
\end{align}
Here $m_N^*$ is the density dependent effective nucleon mass. In our model consideration we suppose $m_N^*=m_N$.

For a finite system the energy per particle (\ref{weise-exp-2}) should be supplemented by the surface energy term, which can be parameterized as~\cite{Tatsumi:2002dq-1,Tatsumi:2002dq-2,Esurf}
\begin{align}\label{Es}
\mathcal{E}_{N,\rm s}(n)=\frac{C_{\rm s}}{A^{1/3}}\frac{\mathcal{E}_{N,\rm v}(n)}{\mathcal{E}_{N,\rm v}(n_0)}\Big(\frac{n}{n_0}\Big)^{4/3}\,,
\end{align}
for densities $n\lsim 1.5\,n_0$, yielding the  surface contributions to the incompressibility, $K_{\rm s}$, $K_N=K_{\rm v}+K_{\rm s}$, and to the Landau parameter, $f_{00,\rm s}$, $f_{00}=f_{00,\rm v}+f_{00,\rm s}$. Below we use $C_{\rm s}=11$\,MeV and  $A=125$. Then the  nucleon energy
\begin{align}
\mathcal{E}_{N}(n)=\mathcal{E}_{N,\rm v}(n)+\mathcal{E}_{N,\rm s}(n)
\label{ENs}
\end{align}
has the minimum at $\tilde{n}_0 \approx 0.9 n_0 \approx 0.15$\,fm$^{-3}$. The  energy per particle in the minimum is $\mathcal{E}_{N}(\tilde{n}_0)= -13.9$\,MeV.  The  Landau parameter
\begin{align}
f_{00}(n) &=f_{00,\rm v}(n)+ \frac{C_{\rm s}}{A^{1/3}}
\frac{p_\rmF^2/m_N}{\mathcal{E}_{N,\rm v}(n_0)} \Big(\frac{n}{n_0}\Big)^{4/3}
\nonumber\\
&\times\Big(\frac{9}{5} - c_1\frac{70 p_\rmF}{9 m_N}
+ c_2\frac{ 88 p_\rmF^2}{9 m_N^2} + c_3\frac{12 p_\rmF^3}{m_N^3}\Big)\,.
\label{f0_Ns}
\end{align}
The parameter $f_{00}$ is equal to $-1$ for two densities
\begin{align}
n_1=0.4216\times 10^{-2}\,n_0, \,\, n_2= 0.5778\,n_0,
\end{align}
and $f_{00}(n)<-1$  in the density region $n_1<n<n_2$\,. Thus, in this density range the formation of the scalar condensate is possible.

\begin{figure}
\centerline{\includegraphics[width=7cm]{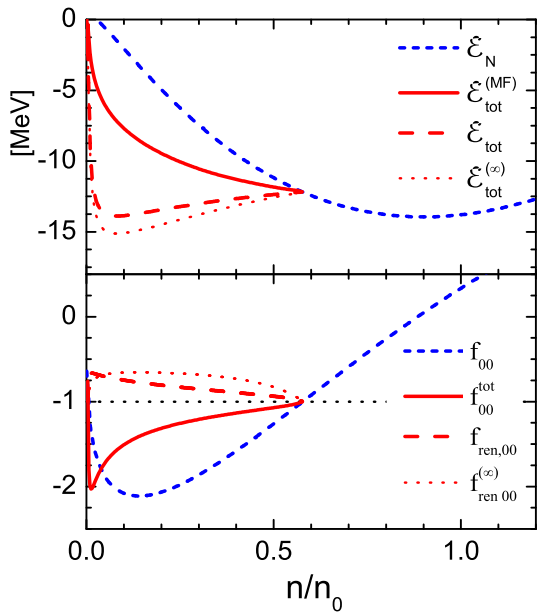} }
\caption{(Color online) Energy per particle (upper panel) and the scalar Landau parameter (lower panel) as functions of the nucleon density in the isospin-symmetric nuclear matter. Short-dash lines show $\mathcal{E}_N$ and $f_{00}$ calculated with Eqs.~(\ref{ENs}) and (\ref{weise-exp-2}), (\ref{Es}), and
Eqs.~(\ref{f0_N}), (\ref{f0_Ns}), respectively, not including the scalar $\phi$ condensate. Solid lines show $\mathcal{E}_{\rm tot}^{\rm MF}$ and $f_{00}^{\rm tot}$ which include the scalar condensate on the mean-field level and are calculated with the solution (\ref{f00-K285}) of the self-consistency equations (\ref{diff-ftot-2}) and (\ref{integ-ftot}).
Long-dashed lines correspond to the energy ${\mathcal{E}}_{\rm tot}(n)$ and
the parameter $f_{\rm ren,00}$, which take into account contributions from the overcondensate excitations, and are determined by Eqs.~(\ref{Erec}) and (\ref{Ifun}) and by Eq.~(\ref{fren}) with $f_{00}$ given by (\ref{f0_Ns}).
Dotted lines are the energy ${\mathcal{E}}_{\rm tot}^{(\infty)}(n)$ and the parameter $f_{\rm ren,00}^{(\infty)}$ calculated in the limit $\Lambda\to \infty$, see Eq.~(\ref{fren-infty}).
} \label{fig:Nucl-cond}
\end{figure}

In the given model we deal with a homogeneous condensate. Then in the mean-field approximation the energy per particle gets additional contributions owing to the condensate, given  by Eq.~(\ref{FB}) for $T=0$:
\begin{align}
\mathcal{E}^{(\rm MF)}_{\rm tot}(n)&=\mathcal{E}_{N}(n) + \mathcal{E}_{\rm B}(n;f_{00}^{\rm tot}(n)),
\label{Etot}\\
\mathcal{E}_{\rm B}(n,f)  &
=-6\epsilon_{\rm F}(n){(f+1)^2}/{f^2} \,.
\label{Eb-f}
\end{align}
The density dependence of the parameter $f_{00}^{\rm tot}$ follows from the solution of the differential equation, cf. Eq.~(\ref{diff-ftot}),
\begin{align}
f_{00}^{\rm tot}(n)=f_{00}(n)+\frac{3 n}{2\epsilon_{\rm F}}
\frac{\rmd^2}{\rmd n^2}[n\mathcal{E}_{\rm B}(n,f_{00}^{\rm tot}(n))]\,,
\label{diff-ftot-2}
\end{align}
where  $f_{00}(n)$ is given by Eqs.~(\ref{f0_N}) and (\ref{f0_Ns}). We assume that the phase transition to the state of the scalar condensate $\phi$ is of the second order, therefore, at the points where $f_{00}=-1$ there should be $f_{00}^{\rm tot}=-1$. Iterative procedure for solving the differential equation (\ref{diff-ftot-2}) badly converges since, as we see below, the solution is rapidly varying in the narrow interval of densities. So, we use a variational method constructing a trial function which would satisfy simultaneously Eq.~(\ref{diff-ftot-2}) and the corresponding integral equation
\begin{align}
\mathcal{E}_{N}(n) +\mathcal{E}_{\rm B}(n;f_{00}^{\rm tot}(n))=I[f_{00}^{\rm tot}](n),
\label{integ-ftot}
\end{align}
where we introduced the functional
\begin{align}
I[f](n,C,\widetilde{C})&=\frac{1}{n}\intop_0^n\rmd n'\intop_0^{n'}\rmd n''
\frac{2\epsilon_{\rm F}(n'')}{3 n''} (f(n'')+1)
\nonumber\\
&
+C+\widetilde{C}/n\,
\label{Ifun}
\end{align}
with $C$, $\widetilde{C}$ as integration constants.
We chose a trial solution in the form
\begin{align}
f_{00}^{\rm tot}(n)=\sum_{i=1}^4 b_i \, u^{\beta_i}\,,\quad u=n/n_0\,.
\label{f00-trial}
\end{align}
Two parameters $b_1$, $b_2$ and the parameter $C$  are fixed by the conditions $$f_{00}^{\rm tot}(n_1)=-1=f_{00}^{\rm tot}(n_2)\,,\,\,\,I[f_{00}^{\rm tot}](n_2,C,\widetilde{C})=\mathcal{E}_{\rm N}(n_2).$$ Other six parameters in (\ref{f00-trial}) and $\widetilde{C}$ are found from the requirement of the best fulfillment of Eqs.~(\ref{integ-ftot}) and (\ref{diff-ftot-2}), i.e. by the minimization of the expression
\begin{align}
\sum_i\{ [I[f_{00}^{\rm tot}](n,C,\widetilde{C})-\mathcal{E}_{N}(n)-\mathcal{E}_{\rm B}(n;f_{00}^{\rm tot}(n))]^2 m_N^{-2}
\nonumber\\
+[f_{00}^{\rm tot}(n)-f_{00}(n)-\frac{3 n}{2\epsilon_{\rm F}}\frac{d^2 n\mathcal{E}_{\rm B}(n,f_{00}^{\rm tot})}{d n^2}]^2\}|_{n=n_i},
\nonumber \end{align}
where the sum is taken for ten density values equally spaced within the interval from 0.05 to $0.55\,n_0$.
For the found solution
\begin{align}
f_{00}^{\rm tot}(n) &=
\frac{1.157\times 10^{-4} }{u^{1.772}}-\frac{0.9531}{u^{0.2014}}
\nonumber\\
&+0.2541 u^{3.449}+ 1.398 (3u/2)^{27.88}
\label{f00-K285}
\end{align}
with $C= -0.3880\times 10^{-2}$ and $\widetilde{C} = 0$
the relative discrepancy between the left and right sides of Eqs.~(\ref{integ-ftot}) and (\ref{diff-ftot}) does not exceed 2.5\% for densities $0.1\,n_0\lsim n\lsim n_2$.
The solution (\ref{f00-K285}) is shown in Fig.~\ref{fig:Nucl-cond}\,(lower panel) by solid line together with  the original interaction parameter $f_{00}(n)$ presented by short-dashed line. We see that $f_{00}^{\rm tot}$ is much closer to the critical value $-1$ for $0.1n_0\lsim n\lsim n_2$ than  $f_{00}(n)$.

The energy per particle  calculated in the mean-field approximation folllowing Eq.~(\ref{Etot}), $\mathcal{E}^{(\rm MF)}_{\rm tot}(n)$, is shown in the upper panel in Fig.~\ref{fig:Nucl-cond} by solid line and the purely nucleon energy term, $\mathcal{E}_{N}(n)$ given by Eq. (\ref{ENs}), by short-dashed line. The inclusion of the condensate reduces the energy considerably. The equation of state is softened thereby. However $\mathcal{E}^{(\rm MF)}_{\rm tot}(n)$ has only one (global) minimum for $n=\tilde{n}_0$. Therefore, if the mean field approximation were valid, the condensate might start to develop in the course of the collapse of the system to the global minimum at $n=\tilde{n}_0$ but finally it would disappear.

Beyond the mean-field approximation the total energy density includes  contributions from the overcondensate excitations,
${E}_{\rm tot}(n)= {E}_{\rm tot}^{\rm (MF)}(n) + {E}^{\rm (fluc)}(n)$. As discussed in Sect.~\ref{sec:Cond-rec}, taking quadratic fluctuations  into account, ${E}_{\rm tot}(n)$ is reconstructed according to Eq.~(\ref{frenE}), with the renormalized Landau parameter $f_{\rm ren,00}$ calculated from Eq.~(\ref{fren}) with the amplitude $f^{\rm tot}_{00}(n)$ given in Eq.~(\ref{f00-K285}).
The renormalized amplitude $f_{\rm ren,00}$ is shown in the lower panel in Fig.~\ref{fig:Nucl-cond} by long-dashed line. Since $f_{\rm ren,00}(n)>-1$ for $n_1 <n<n_2$, the Pomeranchuk instability disappears and  zero- and first-sound modes prove to be stable in presence of the scalar $\phi$ condensate being calculated with account for quadratic fluctuations.

The reconstructed energy per particle can be expressed through the integral (\ref{Ifun}) as
\begin{align}
\mathcal{E}_{\rm tot}(n) = I[f_{\rm ren,00}](n,C',\widetilde{C}')
\label{Erec}
\end{align}
with the parameters $C'$ are $\widetilde{C}'$ found from the continuity conditions
$\mathcal{E}_{\rm tot}(n_{1,2})=\mathcal{E}_{N}(n_{1,2})$. The result is shown by long-dashed line in Fig.~\ref{fig:Nucl-cond}\,(upper panel). Because of the appearance of the condensate and the reconstruction of the nucleon interaction, the energy decreases even more now, so that a metastable minimum develops, at density $n_{\rm ms}=0.081\,n_0$ in our example with the value $-13.8$\,MeV being 100\,keV above  the minimum value at the equilibrium density $\tilde{n}_0$.

For comparison we also consider the limit of an artificially increased $\phi^4$ self-repulsion.  In the limiting case $\Lambda\to \infty$
\begin{align}
\mathcal{E}_{\rm B} =0,\,\, f_{00}^{\rm tot}=f_{00}(n), \,\, f_{\rm ren, 00}^{(\infty)}=-\frac{f_{00}}{2\,f_{00}+1}\,,
\label{fren-infty}
\end{align}
and the energy per particle is calculated as
\begin{align}
\mathcal{E}_{\rm tot}^{\rm (\infty)}(n)=I[f_{\rm ren,00}^{(\infty)}](n,C'',\widetilde{C}'')
\end{align}
with $C''$ and $\widetilde{C}''$ following from the matching of $\mathcal{E}_{\rm tot}^{\rm (\infty)}$ with $\mathcal{E}_N$ at  densities $n=n_{1,2}$.
As we see in Fig.~\ref{fig:Nucl-cond}\,(lower panel, dotted line) $f_{\rm ren, 00}^{(\infty)}>f_{\rm ren, 00}$ almost for all densities $n<n_1$. Correspondingly, the energy per particle $\mathcal{E}_{\rm tot}^{\rm (\infty)}$ shown by dotted line in the upper panel in Fig.~\ref{fig:Nucl-cond} becomes even smaller than $\mathcal{E}_{\rm tot}$ and the absolute minimum of the energy per particle $\mathcal{E}_{\rm tot}^{\rm (\infty)}$ equal $-15.12$\,MeV is located at $n=0.081\,n_0$. We presented this result only to show a possible range of variation of $\mathcal{E}_{\rm tot}$. In reality the limit $\Lambda\to \infty$ is not reached since  it is definitely not legitimate to restrict by quadratic fluctuations if $\Lambda\gg 1$.

The following remark is in order. To demonstrate a possibility of a meta-stable state in the nuclear matter at a subsaturation density owing to the formation of the $\phi$ condensate, we started from a phenomenological parametrization of the nucleon contribution to the energy (\ref{weise-exp-2}), which does not include contribution of the zero-sound-like excitations. We could use  other model, e.g. the relativistic mean-field (RMF) Walecka model or any of its extensions. Within the RMF models for all $n$ there exists a $\sigma$ mean field directly associated with the nucleon scalar density, and in the isospin-symmetric nuclear matter in some density interval, $n_1 <n<n_2$, one has $f_0 (n)<-1$, cf.~\cite{Matsui,Maslov:2015wba}.
If now we take the nucleon system at $T=0$, fixed baryon number and a large volume $V$, having initially a density $n$, $n_1<{n}<n_2$, it will collapse to the ground state with $n\simeq {n}_0$ filling the corresponding part of the volume. In this state we have $f_0 (n_0)>-1$, $\sigma =\sigma (n_0)\neq 0$, and excitation modes are stable. Although $\sigma (n)$ can be considered formally as a scalar field condensate, there is no new physics in such a ``$\sigma$ condensation", as has been argued by Baym long time ago in Ref.~\cite{Baym:1973zk}. The same physics can be described also in any other model, e.g. in the chiral perturbation approach of Ref.~\cite{KFW02} or variational calculations~\cite{APR}, without explicit introduction of any scalar field.
Our scalar condensate field $\phi_0$ does not appear in an RMF approximation, since in the mean-field approximation zero-sounds do not arise. However, if we extend the RMF model by inclusion of contributions from the zero-sound-like excitations, for densities $n_1<{n}<n_2$ we would repeat the calculations performed above with the only replacement of the energy $\mathcal{E}_{N,\rm v}$ given by Eq.~(\ref{weise-exp-2}) by the corresponding quantity calculated within the chosen RMF model. Then the scalar $\phi$ condensate appears as the result a condensation of unstable excitations on top of the $\sigma(n)$ mean field.  We must stress at this point that the $\phi$ field is not reduced to $\sigma (n)$ computed in RMF approximation and appears only in the density interval $n_1 <n<n_2$, where $f_0 (n)<-1$. Moreover, as we have shown, for the appropriate momentum-dependence of the Landau parameter the $\phi$ condensate occurs in $k\neq 0$ state whereas the $\sigma$ mean field in the RMF models corresponds to $k=0$.

Dynamically the $\phi$ condensate evolves  in parallel with growing of the unstable zero- and first-sound modes. Using the estimation \cite{Skokov:2008zp-1,Skokov:2008zp-2} of the coefficient $c\sim 6\cdot 10^{-3}\,$fm$^2$ performed for $n\simeq 0.4 n_0$, and for the surface tension $\sigma (T=0)\simeq 4$ MeV$/$fm$^2$, from Eqs.~(\ref{hk}) and (\ref{homogk}) we see that the $\phi$ condensate amplitude  grows typically at the same rate as the amplitude of the first-sound modes.

\section{Condensate of Bose excitations  with non-zero momentum
in a moving Fermi liquid}\label{sec:moving}

 Let us apply the constructed above formalism to the analysis of
a possibility of the condensation of zero-sound-like excitations with a
non-zero momentum and frequency in a moving Fermi liquid.  The
main idea was formulated in
Refs.~\cite{Pitaev84,V93,Vexp95,Melnikovsky,BP12,KV2015,Kolomeitsev:2015dua}.
When a medium moves rectilinearly with the velocity $u>u_{\rm
L}={\rm min}[\om (k)/k]$ (where the minimum is realized at
$k=k_0\neq 0$),  it may become energetically favorable to transfer
a part of the momentum from the particles of the moving medium to
a condensate of collective Bose excitations with the momentum
$k_0$. The condensation may occur, if in the spectrum branch $\om
(k)$ there is a region with a small energy at sufficiently large
momenta.

As in Ref.~\cite{V93}, we consider a fluid element of the medium
with the mass density $\rho$ moving with a non-relativistic
constant velocity $\vec{u}$. The quasiparticle energy $\om (k)$ in
the rest frame of the fluid is determined from the dispersion
relation
\begin{align}
\label{disp} \Re D_{\phi}^{-1}(\om, k)=0\,.
\end{align}
We continue to exploit the complex scalar condensate field
described by the simplest running-wave probing function, cf.
Eq.~(\ref{running}), and the Lagrangian density (\ref{Lagr}), but now
for the condensate of excitations.

The appearance of the condensate with a finite momentum
$\vec{k}_0$, frequency $\om =\om (k_0)$ and an amplitude
$\phi_0$ leads to a change of the fluid velocity from $\vec{u}$
to $\vec{u}_{\rm fin}$, as it is required by the momentum
conservation
\begin{align}
\label{momentumcons} \rho \vec{u}
=\rho \vec{u}_{\rm fin} +\vec{k}_0 Z^{-1}_0 |\phi_0^2| \,,
\end{align}
where $\vec{k}_0 Z^{-1}_0 |\phi_0^2|$ is the density of the momentum
of the condensate of the boson quasiparticles with the
quasiparticle weight
\begin{align} Z^{-1}_0 (k_0) = \Big[
\frac{\partial}{\partial \om}\Re D^{-1}_{\phi}(\om,
k)\Big]_{\om(k_0),k_0}>0\,. \label{Zinv}
\end{align}

If in the absence of the condensate of excitations the energy
density of the liquid element was $E_{\rm in}=\rho u^2/2$, then in
the presence of the condensate of excitations, which takes a part
of the momentum, the energy density becomes
\begin{align}
\label{Ef} E_{\rm fin}=\half\rho u_{\rm fin}^2 +\om (k_0)Z^{-1}_0
|\phi_0|^2 +\half \Lambda (\om (k_0), k_0) |\phi_0|^4.
\end{align}
Here the last two terms appear because of the classical field of the condensate of  excitations.
The gain in the energy density due to the condensation, $\delta E=E_{\rm fin} -E_{\rm in}$, is equal to
\begin{align}
\delta E = -[\vec{u}\vec{k}_0 -\om(k_0)]\,Z^{-1}_0(k_0)\,
|\phi_0|^2 +\half\widetilde{\Lambda}|\phi_0|^4\,,
\label{dE}
\end{align}
where
\begin{align}
\label{lambdatilde} \widetilde{\Lambda}=\Lambda (\om
(k_0),k_0)+(Z^{-1}_0(k_0))^2 k_0^2/\rho\,.
\end{align}
For $\om =0$, $\Lambda(0, k_0)$ is calculated explicitly, cf.
Eq.~(\ref{lamb-c}). Note that above equations hold also for
$\Lambda =0$.

The condensate of excitations is generated for the velocity of the
medium exceeding the Landau  critical velocity, $u>u_{\rm L} =\om
(k_0)/k_0$, where the direction of the condensate vector
$\vec{k}_0$ coincides with the direction of the velocity,
$\vec{k_0}\parallel \vec{u}$, and the magnitude $k_0$ is
determined by the equation $\om(k_0)/k_0 =\rmd\om(k_0)/\rmd k$.
The gain in the energy density after the formation of the
classical condensate field with the amplitude $\phi_0$ and the
momentum $k_0$ is then
\begin{align}\label{condE}
\delta E =- Z^{-1}_0(k_0) [u k_0 -\om (k_0)]\phi_0^2 + \half\widetilde{\Lambda} \phi_0^4\,.
\end{align}
The amplitude of the condensate field is found by minimization of the energy. From (\ref{condE}) one gets
\begin{align}
\label{varphi}
\phi_0^2 =Z^{-1}_0 (k_0)\frac{u k_0 -\om (k_0)}{\widetilde{\Lambda}}\theta (u-u_c)\,.
\end{align}
The resulting velocity of the medium becomes
\begin{align}
\label{finvel} u_{\rm fin} =u_c +\frac{(u-u_c)\theta
(u-u_c)}{1+[Z^{-1}_0 (k_0)]^2 k_0^2/(\Lambda (\om
(k_0),k_0)\rho)}\,.
\end{align}
For a small $\Lambda$, we have $u_{\rm fin} =u_c +O(\Lambda)$.

For the repulsive interaction $f_0>0$ there is real zero-sound
branch of excitations $\omega_{\rm s} (k)\approx k v_\rmF (s_0 +
s_2\, x^2+s_4\,x^4)$, where the parameters $s_i$ depend on the
coupling constants $f_{00}$ and $f_{02}$ according to
Eqs.~(\ref{s0-coeff}), (\ref{s2-coeff}), and (\ref{s4-coeff}). As
shown in Sect.~\ref{sec:spec-f0pos} the ratio $\omega_{\rm s}
(k)/(k v_\rmF)$ has a minimum at $k_{0}=p_\rmF\sqrt{-s_2/(2s_4)}$
provided $f_{02}$ is smaller than $f_{\rm crit,
02}=-f_{00}^2/[12(s_0^2-1)^2]$.
The Landau critical velocity of the medium is  equal to $u_{\rm
L}/v_\rmF=\om_{\rm s}(k_0)/(v_\rmF k_0) \approx s_0-s_2^2/(2s_4)$.
The quasiparticle weight of the zero-sound mode~(\ref{Zinv}) is
\begin{align}
Z_0^{-1}(k_0)=\frac{a^2
N}{k_0v_\rmF}\frac{\partial\Phi(s,x)}{\partial
s}\Big|_{\frac{\om_s(k_0)}{v_\rmF k_0},\frac{k_0}{p_\rmF}}\,.
\label{Z-fact}
\end{align}
The amplitude of the condensate field~(\ref{varphi}) can be written as
\begin{align}
\label{phif} \phi^2_0 =Z_0^{-1}(k_0) k_0\frac{u -u_{\rm L}
}{\widetilde{\Lambda}} \theta(u-u_{\rm L})\, .
\end{align}
  The energy density gained
owing to the condensation of the excitations is
\begin{align}\label{deltaEex}
\delta E=- k_0^2(Z^{-1}_0(k_0))^2\frac{(u -u_{\rm L}
)^2}{2\widetilde{\Lambda}}\theta
 (u-u_{\rm L})\,.
\end{align}
For a small $\lambda$, Eqs.~(\ref{phif}), (\ref{deltaEex}) simplify as
 \begin{align}
\phi^2_0 &=\rho \frac{(u -u_{\rm L})}{Z^{-1}_0(k_0) k_0}\theta
(u-u_{\rm L}), \nonumber\\ \delta E &=-\frac{\rho}{2}(u-u_{\rm
L})^2 \theta (u-u_{\rm L})\,.
 \end{align}

Finally we note that the Landau parameter $f_0(k_0)$ should be recalculated with account for the condensate in the moving Fermi liquid, except the vicinity of the critical point where the corrections are small and can be neglected.

The instability condition for the mode with $\vec{u}\vec{k}_0>\om(k_0)$ which was found at hand of the energy balance in Eq.~(\ref{dE}) can be obtained differently.  If we transfer our description in the laboratory frame we will shift variables as \cite{Vexp95} $\Phi (\omega, k,n)\to \Phi (\omega -\vec{k}\vec{u}, k,n)$. Then from Eqs.~(\ref{Tph-sol}), (\ref{PhiLow}) follows that the imaginary part of $\Phi$ changes its sign for $\omega <\vec{k}\vec{u}$. Thus the instability first arises for modes with $\omega <ku$ at $\vec{k}\parallel\vec{u}$, like for the Cherenkov radiation. Then the excitations with the momentum $k=k_0$ may start to populate the spectral branch $\omega (k)$, when the minimum $\omega (k)$ touches the value $\vec{k}\vec{u}$ (i.e, the condition of the coincidence of the group and phase velocities).

Now let us consider two interpenetrating dilute streams of fermions which are supposed to interact very weakly if at all. This model may
have a number of physical applications -- just to mention a possible manifestation
of pion instabilities and Cherenkov-like radiation in peripheral heavy-ion collisions, cf. \cite{Vexp95,Pirner:1994tt}.

Let us consider a peripheral nucleus-nucleus ($A+A$) collision in the frame associated with one of the nuclei (the target frame). For a momentum of the excitation $k$ there exists a minimal
value of the projectile momentum, $p_{\rm lab}$, above which the product of two momentum Fermi distributions, $n_{\rm F}(\vec{p})$, of target and projectile fermions vanishes $n_{\rm F}(\vec{p}\,)n_{\rm F}(\vec{p}+\vec{p}_{\rm lab}+\vec{k}\,)=0$\,. Then excitations from one Fermi sphere cannot overlap with the ground state distribution in the other
Fermi sphere. This condition is satisfied already for the laboratory energy $E_{\rm lab}\gsim 200\,{\rm MeV}/A$. All the results obtained above do hold after the replacement
$$f_0(n) \Phi (\omega, k,n)\to f_0(n/2)[\Phi (\omega, k,n/2)+\Phi (\omega -\vec{k}\vec{u}, k,n/2)].$$ For example,  for $\vec{k}\perp\vec{u}$ we have $$f_0(n) \Phi (\omega, k,n)\to 2f_0(n/2) \Phi (\omega, k,n/2)$$ and for $\omega =0$, $k\ll 2 p_{\rm F}$ we arrive at the simple replacement  $f_0(n)\Phi (\omega, k,n)\to 2f_0 (n_0/2)$. For a smooth density dependent value $f_0$ it further reduces to $f_0\Phi (0, k, n)\simeq f_0\to 2f_0$. If so, the Pomeranchuk instability would arise already for the values of $f_0 (n)$ larger than $-1$; e.g., in the just considered simplified case of a smooth $f_0(n)$ function the instability would occur $f_0 (n)<-1/2$ rather than for $f_0 (n)<-1$. The instability would provoke a growth of the scalar condensate field $\phi$ with $\vec{k}_0\perp \vec{p}_{\rm lab}$ in the course of peripheral heavy-ion collisions.

\section{Conclusion}\label{sec:conclude}

We described the spectrum of scalar excitations in normal Fermi liquids for various values of the Landau parameter $f_0$ in the particle-hole channel including its
dependence on the momentum exchanged in the particle-hole channel. For $f_0 >0$ we found a condition on the momentum dependence of $f_0(k)$ when the zero-sound excitations with a
non-zero momentum can be produced in the medium moving with the
velocity larger than the Landau critical velocity. Such
excitations will form an inhomogeneous Bose condensate, which properties are studied.  For
$-1<f_0 <0$ there exist only damped excitations.

For $f_0<-1$ we studied the instability of the spectrum with respect to the growth
of the zero-sound-like excitations (Pomeranchuck instability) and the excitations of the first sound. The surface tension coefficient above which the zero-sound-like mode grows more rapidly than the hydrodynamic one (for ideal hydrodynamics) is estimated.

We derived an effective Lagrangian for the zero-sound-like modes $\phi$ with the scalar boson quantum numbers by performing bosonization of a local fermion-fermion interaction.
The exchange of the boson describes a particle-hole interaction in the scalar channel determined by a Landau parameter $f_0$ corrected by the particle-hole loop diagrams.
For $f_0>0$ poles of such a boson exchange amplitude constitute the spectrum of the zero-sound mode (so $\phi$ is the scalar field which spectrum has the zero-sound branch). For $-1<f_0<0$, the same scalar boson field $\phi$ describes a damped mode. For $f_0<-1$, the field $\phi$ corresponds to an unstable growing mode. We found that this instability can be tamed if the Fermi liquid generates a static homogeneous scalar classical field $\phi$. If the parameter $f_0$ is momentum-dependent there may appear an inhomogeneous condensate with $k=k_{\rm c}\neq 0$, if $f_0(k_{c})<f_0^{\rm cr}$, where $f_0^{\rm cr}\leq -1$.
Properties of the $\phi$ condensate state are investigated.

The situation is analogous to the well-known description of a pion condensation in nuclear matter, see Ref.~\cite{Migdal78,MSTV90} for a review. To remind: One may consider only nucleonic degrees of freedom and treat the pion condensation as a periodic redistribution of nucleons (alternating spin layers). Equivalently, one may separate the pion-exchange contribution from the nucleon-nucleon interaction and add the pion-nucleon interaction terms to the free pion term in the Lagrangian. Then, the poles of a dressed pion propagator determine branches of the pion spectrum (pion zero sound is one of them) and the critical density, $n_c$ of a pion-field instability can be identified. For $n>n_c$ one speaks about a static classical field representing the pion condensate. Now the analogy with our case is complete: the pion exchange channel in the nucleon-nucleon interaction $\Longleftrightarrow$ the scalar channel;
the Landau-Migdal parameter $g'_0$ in case of the iso-symmetric nuclear matter $\Longleftrightarrow$ the parameter $f_0$ in our case; the pion zero sound $\Longleftrightarrow$ the scalar zero sound for $f_0>0$; pion  unstable mode ($\omega^2 <0$) for $n>n_c$ $\Longleftrightarrow$ the Pomeranchuk instability for the scalar mode  for $f_0<-1$, $n_1<n<n_2$; formation of a static pion condensate for $n>n_c $ $\Longleftrightarrow$  formation of a static bosonic $\phi$ condensate in the scalar channel for $f_0<-1$.
The important difference of our case is that the Landau parameter $f_0$ must be recalculated after the inclusion of the condensate contribution to the energy density in order to be consistent with the second derivative of the total energy density.

Note that, as we have argued, for $f_0>0$, if the system moves with velocity $u>u_{\rm L}$, there may appear the inhomogeneous $\phi (\omega_c, k_0)$ scalar condensate if the parameter $f_0$ has the appropriate momentum dependence. Such a condensate previously studied in \cite{Pitaev84,V93,Vexp95,BP12} has the same quantum numbers as the static condensate $\phi$  we have discussed   for $f_0<f_0^{\rm cr}$ ($f_0^{\rm cr}=-1$ at $k_{\rm c}=0$).

Our treatment of the condensate of the $\phi$ field arising for  $f_0<f_0^{\rm cr}$ is consistent only if the condensate amplitude is rather small and fluctuations on the top of the condensate yield small corrections to the mean-field terms. This is so only in the vicinity of the critical point, for $|f_0-f_0^{\rm cr}| \ll 1$, when the difference between $|f_0|$ and the renormalized parameter $f_{{\rm ren},0}$, which determines the spectrum of stable excitations on the top of the condensate, is small. We showed that in the presence of the condensate the equation of state softens for $f_0<f_0^{\rm cr}$.  Then we calculated the energy per particle of the isospin-symmetric nuclear matter with inclusion of the homogeneous $\phi$ condensation beyond the validity of the condition  $|f_0| -1\ll 1$ assuming second-order phase transition. This treatment can be considered only as a kind of an estimating trial. We demonstrated a possibility of a novel metastable state at a subsaturation density ($n\sim 0.1 n_0$ for $T=0$ in our example). The presence of this state could potentially manifest itself in the course of  heavy  ion collisions,  in the process of a supernova explosion and, maybe, in highly excited nuclei. Softening of the equation of state at subsaturation nuclear density and a possibility of novel local minimum may influence on the finite size structures in the pasta phase of the neutron star crusts. Since the density at the novel metastable state is much lower $n_0$ and the size of the droplet is large   for a given atomic number $A$, the appearance of such a metastable nuclei may manifest itself via anomalously large cross sections of their interactions. However, the condensate field evolves in parallel with  the growth of the unstable first-sound modes driving the system  to the ordinary liquid-gas mixture determined by the Maxwell construction  or, for very low $T$, to  the ground state with  $n=n_0$. A rough estimate shows that the condensate field  and the growing first-sound modes, typically, develop at the same rate. Therefore, it might be not so simple to observe manifestations of the $\phi$ condensate in mentioned nuclear systems.

Also we argue that in the regime of two interpenetrating dilute fermion streams, as, e.g., in the case of peripheral heavy-ion collisions, the Pomeranchuk instability may develop already for $f_0 > -1$ ($f_0 <-1/2$).

In the future it is required to perform a more careful comparative study of two
possibilities: the dynamical evolution of the Bose condensation for $f_0<-1$ and the ordinary spinodal instability  appearing at the first-order phase transitions in the
Fermi systems for $T\neq 0$. Note that
inclusion of correlations leading to the clustering at low densities should be also taken into account both at zero and non-zero temperatures, see~\cite{Ropke-1,Ropke-2,Ropke-3,Ropke-4,Ropke-5}. Clustering effect might be important  in description of neutron star crusts, warm supernova  EoS and heavy-ion collisions. Not  only $\alpha$
-particles but higher-order clusters should be incorporated. For recent review of this subject see  \cite{Ropke:2014fia-1,Ropke:2014fia-2}. It would be important to search for a possibility of the
realization of the proposed phenomena on examples of Fermi liquids in  condensed matter.  It would be intriguing to study a possibility of the stabilization of the Fermi liquid with  a strong attractive spin-spin interaction, i.e, with the Landau parameter $g_0 <-1$ by a condensate of a virtual-boson field at certain conditions. These questions will be considered
elsewhere.

\begin{acknowledgement}
The work was supported  by the Ministry of Education and Science of the Russian Federation (Basic part), by Slovak Grant No. VEGA-1/0469/15, by ``NewCompStar'' COST Action MP1304. Computing was partially performed in the High Performance Computing Center of the Matej Bel University using the HPC infrastructure acquired in Project ITMS 26230120002 and 26210120002 (Slovak infrastructure for high-performance computing) supported by the Research \& Development Operational Programme funded by the ERDF.
\end{acknowledgement}



\begin{thebibliography}{99}

\bibitem{FL}
L.D.~Landau, Sov. Phys. JETP {\bf 3}, 920 (1956); {\it ibid.} {\bf 5}, 1011
(1957); {\it ibid.} {\bf 8}, 70 (1959).

\bibitem{LP1981}
E.M.~Lifshitz and L.P.~Pitaevskii, {\it Statistical Physics,
Part 2} (Pergamon, 1980).

\bibitem{Nozieres}
D.~Pines and Ph.~Nozieres, {\it The Theory of Quantum Liquids} (W.A. Benjamin, N.Y., 1966).

\bibitem{GP-FL}
G.~Baym and  Ch.~Pethick,
{\it Landau Fermi-liquid Theory}
(Wiley-VCH, Weinheim, 2004, 2nd. Edition).


\bibitem{Mjump-1}
A.B.~Migdal,
Sov.\ Phys.\ JETP 5, 333 (1957).
\bibitem{Mjump-2}
V.M.~Galitsky and A.B.~Migdal,
Sov.\ Phys.\ JETP {\bf 7}, 96 (1958).
\bibitem{Mjump-3}
A.B.~Migdal, Sov.\ Phys.\ JETP {\bf 16}, 1366 (1963).

\bibitem{M67}
A.B.~Migdal, {\it Theory of Finite Fermi Systems and Properties
of Atomic Nuclei} (Wiley and Sons, N.Y., 1967).

\bibitem{M67a} A.B.~Migdal,
{\it Teoria Konechnyh Fermi System i Svoistva Atomnyh Yader}
(Nauka, Moscow, 1983) [in Russian].

\bibitem{Pomeranchuk}
I. Ya.~Pomeranchuk, Sov.\ Phys.\ JETP {\bf 8}, 361 (1958).

\bibitem{pwave-pairing-1}
D.~Fay and A.~Layzer,
Phys. Rev. Lett. {\bf 20}, 187 (1968).

\bibitem{pwave-pairing-2}
M.Yu.~Kagan and A.V~Chubukov,
JETP Lett. {\bf 47}, 614 (1988).

\bibitem{pwave-pairing-3}
M.Yu.~Kagan, {\it Modern Trends
in Superconductivity and Superfluidity} (Springer, Heidelberg,
2013).


\bibitem{Vexp95}
D.N.~Voskresensky,
Phys. Lett. B {\bf 358}, 1 (1995).


\bibitem{Pitaev84}
L.P.~Pitaevskii,
JETP Lett. {\bf 39}, 511 (1984).


\bibitem{V93}
D. N.~Voskresensky,
JETP {\bf 77}, 917 (1993).


\bibitem{Melnikovsky}
L.A.~Melnikovsky,
Phys.\ Rev.\ B {\bf 84}, 024525  (2011).

\bibitem{BP12}
G.~Baym and C.J.~Pethick,
Phys.\ Rev.\ A {\bf 86}, 023602 (2012).

\bibitem{KV2015}
E.E.~Kolomeitsev and D.N.~Voskresensky, Phys.\ Rev.\ C {\bf 91},
025805 (2015).

\bibitem{Kolomeitsev:2015dua}
E.E.~Kolomeitsev and D.N.~Voskresensky,
arXiv: 1501.00731.

\bibitem{RMS}
G.~R\"opke, L.~M\"unchow, and H.~Schulz,
Phys.\ Lett.\ B {\bf 110}, 21 (1982); Nucl.\ Phys.\ A {\bf 379}, 536 (1982).

\bibitem{SVB}
H.~Schulz, D.N.~Voskresensky, and J.~Bondorf,
Phys.\ Lett.\ B {\bf 133}, 141 (1983).

\bibitem{Chomaz:2003dz}
P.~Chomaz, M.~Colonna, and J.~Randrup,
Phys.\ Rept.\  {\bf 389}, 263 (2004).

\bibitem{Margueron:2002wk}
J.~Margueron and P.~Chomaz,
Phys.\ Rev.\ C {\bf 67}, 041602 (2003).

\bibitem{Ravenhall:1983uh}
D.G.~Ravenhall, C.J.~Pethick, and J.R.~Wilson,
Phys.\ Rev.\ Lett.\  {\bf 50}, 2066 (1983).

\bibitem{Maruyama:2005vb}
T.~Maruyama, T.~Tatsumi, D.N.~Voskresensky, T.~Tanigawa, and S.~Chiba,
Phys.\ Rev.\ C {\bf 72}, 015802 (2005).

\bibitem{Migdal78}
A.B.~Migdal, Rev. Mod. Phys. {\bf{50}}, 107 (1978).
\bibitem{MSTV90}
A.B.~Migdal, E.E.~Saperstein, M.A.~Troitsky, and D.N.~Voskresensky, Phys. Rept. {\bf 192}, 179 (1990).

\bibitem{Tatsumi:2002dq-1}
H.~Heiselberg, C.J.~Pethick, and E.F.~Staubo,
Phys.\ Rev.\ Lett.\  {\bf 70} (1993) 1355.

\bibitem{Tatsumi:2002dq-2}
D.N.~Voskresensky, M.~Yasuhira, and T.~Tatsumi,
Nucl.\ Phys.\ A {\bf 723}, 291 (2003).

\bibitem{MTVTEC}
T.~Maruyama, T.~Tatsumi, D.N.~Voskresensky, T.~Tanigawa, T.~Endo, and S.~Chiba,
Phys.\ Rev.\ C {\bf 73}, 035802 (2006).

\bibitem{Voskresensky:1997ub-1}
D.N.~Voskresensky,
Phys.\ Lett.\ B {\bf 392}, 262 (1997).

\bibitem{Voskresensky:1997ub-2}
E.E.~Kolomeitsev and D.N.~Voskresensky,
Nucl.\ Phys.\ A {\bf 759}, 373 (2005).


\bibitem{SaperFayans}
E.E.~Saperstein and S.V.~Tolokonnikov, JETP Lett. {\bf 68}, 553 (1998).


\bibitem{Sadovnikova-1}
V.A.~Sadovnikova and M.G.~Ryskin,
Phys.\ Atom.\ Nucl.\ {\bf 64}, 440 (2001).
\bibitem{Sadovnikova-2}
V.A.~Sadovnikova,
Phys.\ Atom.\ Nucl.\ {\bf 70}, 989 (2007).
\bibitem{Sadovnikova-3}
V.A.~Sadovnikova,
arXiv:1304.0928 (2013).

\bibitem{KS80-1}
S.A.~Fayans, E.E.~Saperstein, and S.V.~Tolokonnikov,  Nucl.\ Phys.\ A {\bf 326}, 463 (1979).
\bibitem{KS80-2}
V.A.~Khodel and E.E.~Saperstein, Nucl.\ Phys.\ A {\bf 348}, 261 (1980).

\bibitem{AAB} A.I.~Akhiezer, I.A.~Akhiezer, and B.~Barts,
Sov.\ Phys.\ JETP {\bf 29}, 1120 (1969).

\bibitem{Backman:1984sx}
S.O.~Backman, G.E.~Brown, and J.A.~Niskanen,
Phys.\ Rept.\  {\bf 124}, 1 (1985).

\bibitem{Voskresensky:1982vd}
D.N.~Voskresensky and I.N.~Mishustin,
Sov.\ J.\ Nucl.\ Phys.\  {\bf 35}, 667 (1982).

\bibitem{Pethick-Ravenhall88}
C.J.~Pethick and D.G.~Ravenhall,
Ann. Phys. {\bf 183}, 131 (1988).

\bibitem{Speth:2014tja}
J.~Speth, S.~Krewald, F.~Gr\"ummer, P.-G.~Reinhard, N.~Lyutorovich, and V.~Tselyaev,
Nucl.\ Phys.\ A {\bf 928}, 17 (2014).

\bibitem{Matsui}
T.~Matsui,
Nucl.\ Phys.\ A {\bf 370}, 365 (1981).

\bibitem{Maslov:2015wba} K.A.~Maslov, E.E.~Kolomeitsev, and D.N.~Voskresensky,
Nucl.\ Phys.\ A {\bf 950}, 64 (2016).




\bibitem{Wambach:1992ik}
J.~Wambach, T.L.~Ainsworth, and D.~Pines,
Nucl.\ Phys.\ A {\bf 555}, 128 (1993).

\bibitem{Skokov:2008zp-1}
V.V.~Skokov and D.~N.~Voskresensky,
JETP Lett. {\bf 90}, 223 (2009).
\bibitem{Skokov:2008zp-2}
V.V.~Skokov and D.N.~Voskresensky,
Nucl.\ Phys.\ A {\bf 828}, 401 (2009).

\bibitem{Randrup10-1}
J.~Randrup,
Phys.\ Rev.\ C {\bf 79}, 054911 (2009).
\bibitem{Randrup10-2}
J.~Randrup,
Phys.\ Rev.\ C {\bf 82}, 034902 (2010).

\bibitem{VS:2010gf-1}
V.V.~Skokov and D.N.~Voskresensky,
Nucl.\ Phys.\ A {\bf 847}, 253 (2010).
\bibitem{VS:2010gf-2}
D.N.~Voskresensky and V.V.~Skokov,
Phys.\ Atom.\ Nucl.\  {\bf 75}, 770 (2012).


\bibitem{Altland-Simons}
A.~Altland and B.~Simons, {\it Condensed Matter Field Theory} (CUP, Cambridge, 2010).

\bibitem{Kopietz} P. Kopietz, e-print arXiv: cond-mat/0605402.


\bibitem{Brovman-1}
E.G.~Brovman and  Yu.~Kagan,
Sov. Phys. JETP {\bf 36}, 1025 (1972).
\bibitem{Brovman-2}
E.G.~Brovman and A.~Kholas,
Sov. Phys. JETP {\bf 39}, 924 (1974).

\bibitem{IKV00}
Yu.B.~Ivanov, J.~Knoll, and D.N.~Voskresensky, Nucl. Phys. A {\bf 672}, 313 (2000).

\bibitem{V84-1}
D.N.~Voskresensky, Phys. Scripta {\bf 29}, 259 (1984).
\bibitem{V84-2}
D.N.~Voskresensky, Phys. Scripta {\bf 47}, 333 (1993).

\bibitem{D82-1}
A.M.~Dyugaev, Sov.\ Phys.\ JETP {\bf 56}, 567 (1982).
\bibitem{D82-2}
A.M.~Dyugaev, Sov.\ J.\ Nucl.\ Phys.\ {\bf 38}, 680 (1983).

\bibitem{Ivanov:2000ma}
Y.B.~Ivanov, J.~Knoll, H.~van~Hees, and D.N.~Voskresensky,
Phys.\ Atom.\ Nucl.\  {\bf 64}, 652 (2001).

\bibitem{Tilly-Tilly}
D.R.~Tilley and J.~Tilley,  {\it Superfluidity and Superconductivity},
IoP Publishing, Bristol, 1990.

\bibitem{Coleman-Weinberg}
S.~Coleman and  E.~Weinberg,
Phys.\ Rev.\ D {\bf 7}, 1888 (1973).

\bibitem{KFW02}
N.~Kaiser, S.~Fritsch, and W.~Weise,
Nucl.\ Phys.\ A {\bf 697}, 255 (2002).

\bibitem{Esurf}
G.~Baym, H.A.~Bethe, and C.J.~Pethick,
Nucl.\ Phys.\ A {\bf 175}, 225 (1971);


\bibitem{Baym:1973zk}
G.~Baym,
Phys.\ Rev.\ Lett.\  {\bf 30}, 1340 (1973).

\bibitem{APR}
A.~Akmal, V.R.~Pandharipande, and D.G.~Ravenhall,
Phys.\ Rev.\ C {\bf 58}, 1804 (1998).

\bibitem{Pirner:1994tt}
H.-J.~Pirner and D.~N.~Voskresensky,
Phys.\ Lett.\ B {\bf 343}, 25 (1995).

\bibitem{Ropke-1}
S.~Typel, G.~R\"opke, T.~Kl\"ahn, D.~Blaschke, and H.H.~Wolter,
Phys.\ Rev.\ C {\bf 81}, 015803 (2010).
\bibitem{Ropke-2}
M.~Hempel and J.~Schaffner-Bielich,
Nucl.\ Phys.\ A {\bf 837}, 210 (2010).
\bibitem{Ropke-3}
M.D.~Voskresenskaya and S.~Typel,
Nucl.\ Phys.\ A {\bf 887}, 42 (2012).
\bibitem{Ropke-4}
F.~Grill, C.~Providencia, and S.S.~Avancini,
Phys.\ Rev.\ C {\bf 85}, 055808 (2012).
\bibitem{Ropke-5}
G.~R\"opke, N.-U.~Bastian, D.~Blaschke, T.~Kl\"ahn, S.~Typel, and H.H.~Wolter,
Nucl.\ Phys.\ A {\bf 897}, 70 (2013).

\bibitem{Ropke:2014fia-1}
G.~R\"opke,
Phys.\ Rev.\ C {\bf 92},  054001 (2015).
\bibitem{Ropke:2014fia-2} 
G.~R\"opke,
Phys.\ Part.\ Nucl.\  {\bf 46},  772 (2015).

\end{thebibliography}
\end{document}